\documentclass[12pt]{article}
\usepackage{amssymb,latexsym}
\usepackage[mathscr]{eucal}
\usepackage{pstricks,pst-node,pst-text,pst-3d}
\newcommand{\ft}[2]{{\textstyle\frac{#1}{#2}}}

\newcommand{\Z}{{\mathbb Z}}
\newcommand{\R}{{\mathbb R}}
\newcommand{\PP}{{\mathbb P}}
\newcommand{\C}{{\mathbb C}}

\newcommand{\beq}{\begin{equation}}
\newcommand{\eeq}{\end{equation}}
\newcommand{\beqa}{\begin{eqnarray}}
\newcommand{\eeqa}{\end{eqnarray}} 
\newcommand{\tr}{\mbox{\rm Tr}} \newcommand{\uno}{\mbox{1
\kern-.59em {\rm l}}}

\newcommand{\tinyyoung}[1]{\mbox{\tiny\young(#1)}}
\usepackage[vcentermath,enableskew]{youngtab}
\newcommand{\nn}{\nonumber}
\begin{document}
%%%%%%%%%%%%%%%%%%%%%%%  FRONTESPIZIO  %%%%%%%%%%%%%%%%%%%%%%%%%%%%%%%%%%%
\begin{titlepage}
\begin{flushright}
{ROM2F/2006/22}\\
\end{flushright}
%%%%%%%%%%%%%%%%%%%%%%%  TITOLO  %%%%%%%%%%%%%%%%%%%%%%%%%%%%%%%%%%%%%%%%%
\begin{center}
{\large \sc  Instanton on toric singularities and black hole countings }\\
\vspace{1.0cm}
%%%%%%%%%%%%%%%%%%%%%%%%%  AUTORI  %%%%%%%%%%%%%%%%%%%%%%%%%%%%%%%%%%%%%%%%%
{\bf Francesco Fucito} and {\bf Jose F. Morales}\\
{\sl Dipartimento di Fisica, Universit\'a di Roma ``Tor Vergata''\\
 I.N.F.N. Sezione di Roma II\\
Via della Ricerca Scientifica, 00133 Roma, Italy}\\
\vspace{0.3 cm}
{\bf Rubik Poghossian}\\
{\sl Yerevan Physics Institute\\
Alikhanian Br. st. 2, 375036 Yerevan, Armenia}\\
\end{center}
\vskip 2.0cm
%%%%%%%%%%%%%%%%%%%%%%%%%  ABSTRACT  %%%%%%%%%%%%%%%%%%%%%%%%%%%%%%%%%%%%%%
\begin{center}
{\large \bf Abstract}
\end{center}
We compute the instanton partition function for ${\cal N}=4$ U(N)
gauge theories living on toric varieties, mainly of type
$\R^4/\Gamma_{p,q}$ including $A_{p-1}$ or $O_{\PP_1}(-p)$
surfaces. The results provide microscopic formulas for the
partition functions of black holes made out of D4-D2-D0 bound
states wrapping four-dimensional toric varieties inside a
Calabi-Yau. The partition function gets contributions from regular
and fractional instantons. Regular instantons are described in
terms of symmetric products of the four-dimensional variety.
Fractional instantons
 are built out of elementary self-dual connections
with no moduli carrying non-trivial fluxes along the exceptional
cycles of the variety. The fractional instanton contribution
agrees with recent results based on 2d SYM analysis. The partition function,
in the large charge limit, reproduces the
supergravity macroscopic formulae for the D4-D2-D0 black hole entropy.

 \vfill
\end{titlepage}
%%%%%%%%%%%%%%%%%%%%%%%%%%%%%%%%%%%%%%%%%%%%%%%%%%%%%%%%%%%%%%%%%%%%%%%%%%%
%%%%%%%%%%%%%%%%%%%%%%%  INIZIO TESTO  %%%%%%%%%%%%%%%%%%%%%%%%%%%%%%%%%%%%
%%%%%%%%%%%%%%%%%%%%%%%%%%%%%%%%%%%%%%%%%%%%%%%%%%%%%%%%%%%%%%%%%%%%%%%%%%%
%\addtolength{\baselineskip}{0.3\baselineskip}
%%%%%%%%%%%%%%%%%%%%%%%%%%%%%%%%%%%%%%%%%%%%%%%%%%%%%%%%%%%%%%%%%%%%%%%%%%%
%%%%%%%%%%%%%%%%%%%%%%%%%%%%%%%%%%%%%%%%%%%%%%%%%%%%%%%%%%%%%%%%%%%%%%%%%%%
%%%%%%%%%%%%%%%%%%%%%%%%%%%%%%%%%%%%%%%%%%%%%%%%%%%%%%%%%%%%%%%%%%%%%%%%%%%
%%%%%%%%%%%%%%%%%%%%%%%%%%%%%%%%%%%%%%%%%%%%%%%%%%%%%%%%%%%%%%%%%%%%%%%%%%%
%%%%%%%%%%%%%%%%%%%%%%%  SECTION 1     %%%%%%%%%%%%%%%%%%%%%%%%%%%%%%%%%%%%
%%%%%%%%%%%%%%%%%%%%%%%%%%%%%%%%%%%%%%%%%%%%%%%%%%%%%%%%%%%%%%%%%%%%%%%%%%%

\tableofcontents

\setcounter{section}{0}

\section{Introduction}

Recent studies in string theory have revealed  remarkable
connections between apparently unrelated subjects like black hole
thermodynamics, topological strings and instanton physics. Not
surprisingly, the key role in the game is played by D-branes. The
typical example being type IIA string on a Calabi-Yau (CY)
threefold. A black hole in the resulting ${\cal N}=2$
four-dimensional supergravity can be built out of D0-D2-D4-D6 bound
states wrapping various cycles inside the CY. The black
hole partition function $Z_{\rm BH}$ can be defined as the thermal
partition function with (canonical) microcanonical ensemble for
(electric) magnetic charges ($Q_0,\vec{Q}_2$) $\vec{Q}_4,Q_6$: \beq
Z_{BH}=\sum_{Q_0,\vec{Q_2}} \Omega(Q_0,\vec{Q}_2,\vec{Q}_4,Q_6)\,
e^{-Q_0 \varphi_0-\vec{Q}_2\cdot \vec{\varphi}_2} \eeq
 with $\Omega(Q_0,\vec{Q}_2,\vec{Q}_4,Q_6)$ counting the number of
 D0-D2-D4-D6 bound states. We will always take $Q_6=0$ (no D6-branes) and write $\vec{Q}_4=N $ to denote
  $N$ D4-branes wrapping a given four-cycle $M$ inside the CY. For large $N$, $Z_{BH}$ can
be computed in supergravity as the exponential of the microcanonical black hole
entropy. The result can be written in the suggestive form \cite{Ooguri:2004zv,Gaiotto:2006ns,Beasley:2006us}
\beq
Z_{\rm BH}=|Z_{\rm top}|^2+\ldots \label{osv} \eeq where $Z_{\rm top}$ is
the partition function of the topological string on the CY and the dots stand for $O(e^{-N})$ corrections.

On the other hand a D0-brane can be thought of as an instanton on the
twisted ${\cal N}=4$ supersymmetric four-dimensional gauge theory (SYM) living on the
D4-brane. In the same spirit, D2-branes are associated to
magnetic fluxes along the D4-worldvolume, i.e. to non-trivial
first Chern classes of the instanton gauge bundle. When the D4-branes wrap an $A_{p-1}$
ALE singularity, the moduli space of the D0-D2-D4 system can be realized by the
ADHM construction \cite{kn}.
 The multiplicities of the D0-D2-D4 bound states are given by the number of
vacua (Witten index) of the gauge theory living at the zero dimensional brane intersection.
This is a quantum mechanics  with target space the ADHM manifold.
Since coordinates and differentials on the ADHM instanton moduli space
can be interpreted as bosonic and fermionic degrees of freedom of the quantum mechanics,
the Witten index gives the Euler number of the instanton moduli space.
The generating function for the Euler character of the instanton moduli spaces
gives the SYM partition function, $Z_{SYM}$, of the ${\cal N}=4$ theory.
This leads to an alternative route to compute the black hole partition function
\beq
Z_{\rm BH}=Z_{SYM}
\eeq
 The relation (\ref{osv}) becomes then a highly non-trivial statement
 about the gauge theory: $Z_{SYM}$ should factorize at large $N$.
This factorization has been confirmed by explicit computations in
\cite{Aganagic:2004js,Aganagic:2005wn,Caporaso:2005ta,Caporaso:2005fp,Caporaso:2006gk}
where the study of the $d=4$ SYM partition function has been addressed via 2d-SYM techniques.
More precisely, the authors study D4-branes wrapping a $O_{\Sigma_g}(-p)$
four-cycle inside the non-compact CY $O(p-2+2 g)\oplus O(-p)\to \Sigma_g$.
The computation was performed in a  $q$-deformed 2d SYM theory following from
dimensional reduction of ${\cal N}=4$ SYM down to $\Sigma_g$. The partition function of the $q$-deformed 2d SYM
 theory was written as a sum over $U(N)$ representations and the factorization was proved to hold
in the large $N$ limit.
After a Poisson resummation the results were cast in a form that resembles a four-dimensional instanton sum.

 In this paper we compute the instanton partition function directly from the four
 dimensional perspective. The study of instantons effects in four-dimensional gauge theories
 has been considered for a long time an "out of reach" task due to the highly non-trivial
 structure of the ADHM instanton moduli spaces. This situation
 drastically changed with the discovery \cite{Nekrasov:2002qd}
 that the instanton partition function (and chiral SYM amplitudes) localize around
 a finite number of critical points in the ADHM manifold. This leads to an impressive
 number of results in the study of multi-instanton corrections to ${\cal N}=1,2,4$ supersymmetric
 gauge theories in $\R^4$ \cite{Flume:2002az,Bruzzo:2002xf,Losev:2003py,Nekrasov:2003rj,Flume:2004rp,Marino:2004cn,Fucito:2004gi,Fucito:2005wc}
 (see \cite{Dorey:2002ik} for a review of multi-instanton techniques before localization and a
complete list of references).

  Here we apply the localization techniques to the study of instantons on toric varieties coming
from blowing up orbifold singularities
  of the type of $M_4=\C^2/\Gamma_{p,q}$. $\Gamma_{p,q}$ denotes a $\Z_p$-action specified by
  the pair of coprime integers $(p,q)$. This includes the $A_{p-1}$ singularities and the blown down
  $O_{\PP_1}(-p)$ surfaces.
The general $(p,q)$ case describes the most general toric singularity in four dimensions.
The ADHM construction of instantons on a $A_{p-1}$-singularity was carried out in \cite{kn}.
In \cite{Fucito:2004ry}  this construction was applied to the study of the
prepotential of ${\cal N}=2$ $SU(N)$ SYM theories on the ALE space (see also \cite{Fucito:2001ha}).
 The results were developed and put in firm mathematical grounds in \cite{Fujii:2005dk}.
Here we generalize these results to the case of D4-D2-D0 bound states on general toric varieties.
We work in details the case of $\C^2/\Gamma_{p,q}$ singularities.
 We start by revisiting the case of instantons on $A_{p-1}$-spaces. We give a complete description
 of the gauge bundle and present an alternative derivation of the instanton partition function which
 naturally extends to instantons on a general (compact or not) toric variety where an explicit ADHM
 construction is not known.
 The resulting instanton formula will be tested against 2d SYM results and supergravity
 macroscopic black hole entropy formulae that suggest that our results apply to
 general (compact or not) toric varities at least in the limit of large instanton charges.
The instanton partition functions
will be written in terms of modular invariant forms consistently
with the $SL(2)$-invariance of the ${\cal N}=4$ gauge theory.

The paper is organized as follows: In Section \ref{sc2}, \ref{scpq} we review and elaborate on the ADHM
 construction of instanton moduli spaces on $\C^2$ and $A_{p-1}$. We give a detail description of the instanton
 Chern classes and compute the SYM partition function.
 Sections \ref{sop} deals with instantons on a general $\C^2/\Gamma_{p,q}$ singularity.
 In section \ref{sbh}, we derive the microscopic D4-D2-D0 black hole
 partition function and test it against supergravity.
 Section \ref{sconcl} summarizes our results. In Appendix
 \ref{fuji} we review the regular/fractional factorization
 algorithm developed in \cite{Fujii:2005dk}.
 In section \ref{storico} we collect
some useful background material in toric geometry.

\section{ADHM on $\C^2$}
\label{sc2}
\setcounter{equation}{0}

The moduli space of self-dual $U(N)$ connections on $\C^2$ can be
described as a $U(k)$ quotient of a hypersurface on $\C^{2 k^2+2 k
N}$ defined by the ADHM constraints
\beqa \label{mommap}
&&[B_1,B_2]+IJ=0\ \,
\nonumber\\
&&[B_1,B_1^\dagger]+[B_2,B_2^\dagger]+II^\dagger-J^\dagger J=\xi\,
{\bf 1}_{k\times k}.\label{adhm} \eeqa The coordinates in $\C^{2
k^2+2 k N}$ are represented by $B_{1,2}$, $I$ and $J$ given by
$[k]\times [k]$, $[k]\times [N]$ and
 $[N]\times [k]$ dimensional matrices respectively.
The $U(k)$ action is defined as
\beq
 B_{\ell} \to U\, B_{\ell}\, U^\dagger  \quad\quad I\to U\, I
\quad\quad J\to J\, U^\dagger
\eeq
 with $U$ a $[k]\times [k]$ matrix of $U(k)$. There is a nice D-brane
description of this system. A $U(N)$ instanton
with instanton number $k$ can be thought as a bound state of $k$ $D(-1)$ and $N$ $D3$-branes. The instanton
moduli  $B_{\ell},I,J$  represent the lowest modes of
open strings connecting the various branes.
The ADHM constraints are identified with the F- and D- term flatness
conditions in the effective 0-dimensional theory.
Finally $\xi$, a Fayet-Iliopoulos term, measures the non-commutativity of spacetime and
is needed in order to regularize the moduli space. All physical quantities we will deal with
are independent of the value of $\xi$ and therefore we can think of it as a regularization artifact.

The tangent space matrices $ \delta B_{1,2},\delta I,\delta J$ can be thought
of as the homomorphisms
\beqa
\delta B_\ell &\in & V\otimes V^*\otimes Q \nn\\
\delta I  &\in & V\otimes W^* \nn\\
\delta J & \in & W\otimes V^*\otimes \Lambda^2 Q \eeqa between the
spaces $V,W,Q$ with dimensions $[k]$, $[N]$ and $[2]$
respectively. In the brane picture $V,W$ parametrize the space of
D(-1) and D3 boundaries while $Q$ is a doublet respect to a
$SU(2)$ subgroup of the Lorentz group. For the sake of simplicity
from now on we will omit the tensor product simbol and use $+$
instead of $\oplus$. Product of spaces will be always understood
as tensor products.

The tangent instanton moduli space can then be written as\cite{nakajima}
 \beqa \label{adhmM} T{\cal M}_k &=& V^*\, V\,
\left[Q-\Lambda^2 Q-1\right] + W^*\, V
+V^*\, W\, \Lambda^2 Q
\eeqa
The contributions with the minus signs come from the three ADHM constraints (\ref{mommap}) and
 the $U(k)$ invariance ($2k^2$ complex degrees of freedom in total).
 The dimension of the moduli space ${\cal M}_k$  can be
 easily read from (\ref{adhmM}) to be ${\rm dim}_{\C} {\cal M}_k=k^2(2-2)+2 k N=2 kN$.

SYM amplitudes involve integrals over the ADHM manifold. For
chiral correlators these integrals localize around isolated fixed
points of the instanton symmetry group $U(k)\times U(N)\times
SO(4)$ on the ADHM manifold \cite{Nekrasov:2002qd}. Moreover for
${\cal N}=4$ SYM the contribution of the vector supermultiplet
cancels against that of the hypermultiplet and the instanton
partition function reduces to the counting of the number of fixed
points, i.e. the Euler character of the instanton moduli space
\cite{Bruzzo:2002xf}.
 In this paper we will mainly deal with the evaluation of this index.

The spaces $V,W$ transform in the fundamental representation of
$U(k)$ and $U(N)$ respectively while $Q$ transforms in the chiral
spin representation of $SO(4)$. We parametrize the action of the
Cartan symmetry group $U(1)^{N+k+2}$ by
$a_\alpha,\phi_s,\epsilon_l$  with $\alpha=1,\ldots,N$,
$s=1,\ldots$ ,$k,l=1,2$. Fixed points are in one to one
correspondence with sets of $N$ Young tableaux $Y=(Y_1,\ldots
Y_N)$ with $k=\sum_\alpha k_\alpha$ boxes distributed between the
$Y_\alpha$'s. The boxes in a $Y_\alpha$ diagram are labelled by
the index $s=(\alpha,i_\alpha,j_\alpha)$ with $i_\alpha,j_\alpha$
denoting the horizontal and vertical position respectively in the
Young diagram $Y_\alpha$. More precisely \beqa \label{solcritical}
e^{i \phi_{s}} &=& e_{\alpha}\, T_1^{-j_\alpha+1}\,
T_2^{-i_\alpha+1} \eeqa with $T_{\ell}=e^{i\epsilon_{\ell}}$,
$e_\alpha=e^{i a_\alpha}$. The tangent instanton moduli space is
spanned by the fluctuations $\delta B_\ell$, $\delta I$, $\delta
J$ satisfying the linearized ADHM constraints around the fixed
point. Chiral SYM amplitudes can be related to the character of the
$U(1)^{k+N+2}$ action evaluated at the fixed points on the tangent space.
 Although we are mainly interested here on the  ${\cal N}=4$ SYM partition given by the blind counting
of fixed points, for future references we will write (when possible) explicit expressions for the
full $U(1)^{k+N+2}$ character.

The character of a space ${\cal H}$ will be defined as follows
\beq
\chi_\epsilon( {\cal H}) \equiv \tr_{\cal H}\, e^{ia_\alpha J^\alpha+i\phi_{s} J^s+i\epsilon_l J^l}\label{chardef}
\eeq
with $J^\alpha,J^s, J^l$ the generators of the Cartan group $U(1)^{k+N+2}$.
In particular at a given fixed point $Y=\{ Y_\alpha \}$ one finds
\beqa
\chi_\epsilon(V)&=& \sum_{(\alpha,i_\alpha,j_\alpha)\in Y_\alpha}\, e_{\alpha}\,T_{1}^{1-j_\alpha}
 T_{2}^{1-i_\alpha} \nonumber\\
\chi_\epsilon( W) &=&\sum_{\alpha =1}^N\,  e_{\alpha}\nn\\
\chi_\epsilon( Q) &=& T_1+T_2;\quad\quad Q=Q_1+Q_2
\label{Vs}
\eeqa
 Plugging (\ref{Vs}) into (\ref{adhmM}) one finds the character
of the instanton moduli space ${\cal M}_Y$ at the fixed point $Y$ \cite{Flume:2002az,Bruzzo:2002xf}
\beq
\chi_\epsilon({\cal M}_Y)= \sum_{\alpha, \beta}^N
{\cal N}^Y_{\alpha,\beta}(T_1,T_2)
\eeq
with
 \beqa \label{trace}
{\cal N}^Y_{\alpha,\beta}(T_1,T_2)&=& e_{\alpha} e_{\beta}^{-1}
\left( \sum_{s_\alpha\in Y_\alpha} \, T_1^{-h_\alpha(s_\alpha)}
T_2^{v_\beta(s_\alpha)+1}+ \sum_{s_\beta\in Y_\beta}\,
T_1^{h_\beta(s_\beta)+1} T_2^{-v_\alpha(s_\beta)}\right) \eeqa and
\beqa
h_\alpha(s_\alpha)=h_\alpha(i_\alpha,j_\alpha)=\nu^{Y_\alpha}_{j_\alpha}-i_\alpha\nn\\
v_\alpha(s_\beta)=v_\alpha(i_\beta,j_\beta)
=\tilde{\nu}^{Y_\alpha}_{i_\beta}-j_\beta \eeqa Here
$\nu^{Y_\alpha}_{i_\alpha} (\tilde{\nu}^{Y_\alpha}_{j_\alpha})$
denotes the lengths of the columns(rows) in the tableau
$Y_\alpha$. In other words $h_\alpha(s_\beta)$
($v_\alpha(s_\beta)$) denotes in general the number of horizontal
(vertical) boxes in the tableau $Y_\alpha$ to the right(on top) of
the box $s_\beta\in Y_\beta$. Notice in particular that
$h_\alpha(s_\beta)$ is negative if $s_\beta$ is outside of the
tableau $Y_\alpha$.

\section{Instantons on $A_{p-1}$ }
\label{scpq}
\setcounter{equation}{0}

In this section we review and elaborate on the ADHM construction of instantons on
a $A_{p-1}$-singularity \cite{kn,Fucito:2004ry,Fujii:2005dk}.
The ALE space $A_{p-1}$ is defined by the quotient $\C^2/\Gamma$ with $\Gamma$ the
$\Z_p$-action with generator
\beq
\Gamma :~~~~  \left(%
\begin{array}{c}
  z_1 \\
  z_2 \\
\end{array}%
\right) \rightarrow  \left(%
\begin{array}{cc}
  e^{2\pi i /p} & 0 \\
  0 &   e^{-2\pi i/p}\\
\end{array}%
\right)\left(%
\begin{array}{c}
  z_1 \\
  z_2 \\
\end{array}%
\right)
\eeq
The moduli space of instantons on $\C^2/\Gamma$ can be found from that on $\C^2$
by projecting onto its  $\Gamma$-invariant component.
In this projection, the fixed points under the action of $U(1)^{N+k+2}$ on the
$\C^2$-ADHM moduli space are preserved since $\Gamma\in U(1)^{N+k+2}$.
However, the projection splits the $\C^2$ instanton moduli into several disjoint pieces classified by the
choice of the $\Z_p$ representations under which the D(-1) instantons and D3-branes transform.
 Fixed points are now described by p-coloured Young tableaux.
More precisely, $U(N)$ instantons
on $\C^2/\Gamma$ are specified by the $N$ sets $\{ (Y_\alpha, r_\alpha) \}$ with $Y_\alpha$ a
tableau with p types of boxes and $r_\alpha$ an integer $mod\, p$.
The label $r_\alpha$ specifies the $\Z_p$ representation $R_{r_\alpha}$ under
which the first box in $Y_\alpha$ transforms. The p choices correspond to the p types of fractional
D3-branes.
More precisely the $r_\alpha$'s specify the embedding of $\Gamma$ into the
$U(1)^{N+2}$ symmetry group
\beq \Gamma:\quad\quad
R_a\to e^{2\pi i a/p} R_a\quad\quad T_1\to e^{2\pi i \over p}\, T_1 \quad\quad T_2\to
e^{-{2\pi i \over p}}\, T_2 \quad\quad
e_\alpha\to e^{2\pi i r_\alpha \over  p}\, e_\alpha
\label{zn} \eeq
 Notice that this action induces
also an action on $V$ via (\ref{solcritical}). Indeed
the box $(i_\alpha,j_\alpha)$ of the tableaux $Y_\alpha$ transforms in the
representation $R_{r_\alpha+i_\alpha -j_\alpha}$.
The spaces $V,W$ decompose as
\beqa
 V&=&\sum_{a=0}^{p-1}\, V_a \, R_a; \quad\quad {\rm dim}V_a=k_a\nn\\
 W&=&\sum_{a=0}^{p-1}\, W_a \, R_a; \quad\quad {\rm dim}W_a=N_a
\eeqa
 with $k_a,N_a$ counting the number of times
the $a^{\rm th}$-representation appears
in $V$ and $W$ respectively. Notice that $N_a$ is specified by the choice of
 $r_\alpha$ via $N_a=\sum_\alpha \delta_{r_\alpha,a}$.
 The unbroken symmetry  group is
 $SU(2)\times U(1)\times\prod_{a=0}^{p-1} U(N_a)\times U(k_a)$ with
  $SU(2)\times U(1)$ the isometry of the ALE space.

  The tangent of the instanton moduli space is then given by the $\Gamma$-invariant
component of the $\C^2$ result\cite{Fucito:2004ry}
\beqa \label{adhmMzp} T{\cal M}_Y &=& \left(  V^*\, V\,
\left[Q-\Lambda^2 Q-1\right] +  W^*\, V
+ V^*\, W\, \Lambda^2 Q \right)^{\Gamma}\\
&=& \sum_a \left(V_a\, \left[V^*_{a+1}\, Q_1+V^*_{a-1}\, Q_2-V^*_{a}
- V^*_{a}\, Q_1\, Q_2
\right]+V_a W^*_a+V^*_{a}\, W_{a}\, Q_1 Q_2 \right)\nn
\eeqa
Here and below the subscript $a$ will be always understood $mod~p$.
 The dimension of the moduli space is given by
\beqa {\rm dim}_{\C} {\cal M}_Y &=&
 \left(k_a\, k_{a+1}+k_a k_{a-1}-2 k_a^2
+2 k_a\, N_a\right)\nn\\
&=& - \,\widehat{C}_{a b}\, k_a\, k_{b}+ 2 k_a\, N_a \label{dimform}
\eeqa
 with $\widehat{C}_{ab}=2 \delta_{ab}-\delta_{a,b+1}
 -\delta_{a,b-1}$ the extended $A_{p-1}$ Cartan matrix.
Repeated indices are understood to be summed over $a,b=0,\ldots,p-1$.
Notice that the complex dimension is always
even, in agreement with the fact that
the instanton moduli space on $A_{p-1}$ is hyperk\"ahler.

The character is given by the $\Gamma$ invariant components of (\ref{trace})
\beq
\chi_\epsilon({\cal M}_Y)= \sum_{\alpha, \beta}^N
{\cal N}^Y_{\alpha,\beta}(T_1,T_2)^\Gamma  \label{charinv}
\eeq
 with ${\cal N}^Y_{\alpha,\beta}(T_1,T_2)^\Gamma$
the
restriction of the $\C^2$ result to those monomials invariant
under (\ref{zn}).

Now let us describe the instanton gauge bundle.
The $A_{p-1}$ singularity can be resolved by the blowing up procedure which consists
in replacing the singularity with $p-1$ intersecting spheres, $\PP^1$ (called exceptional divisors).
With respect to the $\R^4$ case, this leads to new self-dual connections with non-trivial fluxes along
the exceptional divisors. The instanton bundle can then be constructed out of elementary $U(1)$
bundles, ${\cal T}^a,\,\,a=0,\ldots,p-1$, carrying the unit of flux through the exceptional divisors.
 ${\cal T}^0$ denotes the trivial bundle.
Following \cite{gn} we will refer to this bundle as
the tautological bundle.
In the ADHM construction the $U(1)$ bundle ${\cal T}^a$, corresponds to a tableau  with
no boxes $k_1=k_2=\ldots=0$ and $r= a$.

The $p-1$ non-trivial bundles ${\cal T}^a$ with $a=1,\ldots,p-1$  can be used as a basis
 for two forms on $A_{p-1}$ with intersection matrix
 \beq
 \int c_1({\cal T}^a)\wedge c_1({\cal T}^b)=-C^{ab}
  \quad\quad \int_{C_a} c_1({\cal T}^b)=\delta_a^b
 \eeq
 $C^{ab}$, being the inverse of the $A_{p-1}$-Cartan matrix.
Here and below, for the sake of simplicity, we have extended the range of the $a$-index to $a=0,\ldots,p-1$ by
defining $C^{0a}=0$.
The gauge bundle $F_Y$ is given by \cite{kn}
\beqa
F_Y &=&\left(  V^*\,{\cal T}\,
\left[Q-\Lambda^2 Q-1\right] +  W^*\, {\cal T}\right)^{\Gamma}\nn\\
&=& \sum_a {\cal T}^a\, \left(  V^*_{a+1} Q_1+V^*_{a-1} Q_2
-V^*_{a} \,Q_1\, Q_2 -V^*_{a}+W^*_a\right)
\label{gaugeE}
\eeqa
with ${\cal T}=\sum_a {\cal T}^a R_a$ the tautological bundle.

  The Chern characters are given by\footnote{Our conventions:
  $c(F)=\sum_i c_i(F)\, t^i={\rm det}\left(1+t\,{i\,F\over 2 \pi}\right)$,
${\rm ch}(F)=\sum_i {\rm ch}_i(F)\, t^i={\rm tr}\,e^{t\, {i\, F\over 2 \pi}}$,
  $F=dA$.}
  \beqa
  {\rm ch}_1(F_Y) &=&\sum_a \,u_a \,{\rm ch}_1({\cal T}^a) \nn\\
{\rm ch}_2(F_Y) &=& \sum_{a} u_a {\rm ch}_2({\cal T}^a)-{K\over p}\, \Omega\quad\quad~~~~ K=\sum_a k_a
\label{chern}
  \eeqa
with $\Omega$ the normalized volume form of the manifold,
${\rm ch}_1=c_1$, ${\rm ch}_2=-c_2+\ft12 c_1^2$ and
\beq
u_a = N_a+k_{a+1}+k_{a-1}-2\, k_a=N_a-\widehat{C}_{a b}\, k_{b}
\label{ut}
\eeq
Notice that $\sum_a u_a=N$ therefore $u_a$'s is characterized by $p-1$ independent components.

The instanton number $k\in \ft{1}{2p} \Z$ is defined as
  \beqa
k=-\int_M {\rm ch}_2(F_Y) =
\ft12\,\sum_{a} \, C^{aa}\, u_a+{K\over p}= k_0+\ft12\, \sum_a C^{aa} \,N_a\label{c2k0}
\label{chern2}
  \eeqa
with
\beq
C^{aa}={1\over p}(p-a)a
\eeq
 Formula (\ref{c2k0}) shows that the instanton number can be computed
 by counting the numbers of $0$'s in the
Young tableaux set $Y$.

The instanton bundle will be labelled then by $u_a$ with $a=1,2,\ldots,p-1$ and $k$ specifying
the first and second Chern characters respectively.
We will compute the following index
\beq
 {\cal Z}(q,z_a)\equiv \left\langle e^{ \varphi_{2a}\,\int c_1(F)\wedge c_1({\cal T}^a)} \right\rangle
 =\sum_{k,u_a} \chi({\cal M}_{k,u_a}) \, q^k\, e^{-z^a\, u_a} \quad\quad z^a\equiv C^{ab}\, \varphi_{2b}\label{partfunct}
\eeq
where $q=e^{2\pi i\, \tau}$, $\tau$ is the complexified coupling constant and
$\chi({\cal M}_{k,u_a})$ is the Euler number of the instanton moduli space
with first and second Chern characters $u_a$ and $k$ respectively.
 We will refer to ${\cal Z}(q,z_a)$ as the instanton partition function. This index counts the number
of bound states formed by $N$ D4 branes, $u_a$ D2 branes (wrapping the $a$ exceptional divisor) and $k$
D0 branes.
 As we explained before $\chi({\cal M}_{k,u_a})$ can be computed by simply counting
 the number of Young tableaux for a given $k,u_a$.

The result (\ref{partfunct}) can be refined by introducing
the Poincar\'{e} polynomial generating function
\beq
{\cal P}(t,q,z_a)\equiv
\sum_{u_a,k}  P(t|{\cal M}_{k,u_a}) \,  q^k\, e^{-z^a\, u_a}
=\sum_{u_a,k} b_i({\cal M}_{k,u_a})\, t^i\, \,  q^k\, e^{-z^a\, u_a}
\eeq
with $P(t|{\cal M}_{k,u_a})$ the Poincar\'{e} polynomial of the instanton moduli space
and $b_i({\cal M}_{k, u_a})$ its Betti numbers.
Notice that one can specify the fixed point either by
$k,u_a$ or by $\{ k_a \}$. As shown in \cite{Fucito:2004ry} each choice of $\{ k_a \}$
and therefore of $k,u_a$ leads to a disconnect piece of the instanton moduli space, i.e.
$b_0({\cal M}_{k,u_a})=1$.

 The Betti numbers can be computed by taking
\beq
T_1>> e_1>> e_2 >>\ldots   e_N>> T_2 \label{range}
\eeq
 and counting the number of negative eigenvalues in the character
$\chi_\epsilon({\cal M}_{Y})$.
More precisely each fixed point can be associated to a harmonic form on the moduli space
and therefore contributes once to the Poincar\'{e} polynomial.
This contribution is given by $t^{2 n^Y_-}$, with $n_-^Y$ the number
of negative eigenvalues in $\chi_{\epsilon}({\cal M}_{Y})$\footnote{We have checked that the
Poincar\'{e} polynomial coming from (\ref{charinv}) agrees against the results
of \cite{Fucito:2004ry,hausel}.}.

Summarizing, fixed points in the moduli space of $U(N)$
instantons on $\C^2/\Gamma$ are specified by the $N$ sets
$(Y_\alpha,r_\alpha)$, $\alpha=1,\ldots,N$ with $Y_\alpha$
p-colored Young
tableaux and $r_\alpha=0,\ldots,p-1$ an integer specifying the
$\Z_p$-representation under which the first box in
$Y_\alpha$ transforms. Different choices of
$\{ r_\alpha\}$ describe in general different breakings
$\prod_{a=0}^{p-1} U(N_a)\times U(k_a)$ of the
starting $U(N)\times U(K)$ symmetry group.
The tangent space is given by projecting
the $\C^2$ character  under (\ref{zn}).
 The instanton gauge bundle is given by (\ref{gaugeE}) with Chern characters
(\ref{chern}) and (\ref{chern2}). The instanton partition function is defined in (\ref{partfunct})
and can be computed by counting the number of Young tableaux with fixed Chern characteristics.

\subsection{Regular vs fractional instantons}
\label{sap}

For ALE spaces there are two types of instantons: regular and fractional ones.
A regular instanton is an instanton in the regular representation
of $\Gamma=\Z_p$. This type of instanton is free to move (together with
its images) on $\C^2/ \Z_p$. The
moduli space of $K=k\, p$ regular instantons is then given by
choosing symmetrically $k$ points on $\C^2/\Z_p$, i.e.
 ${\cal M}^{\rm reg}_{k p}= (\C^2/\Z_p)^k/S_k \sim
(\C^2 / \Z_p)^{[k]}$, with $M^{[k]}$
the Hilbert scheme of $k$-points on $M$.
  Fractional instantons on the other hand correspond to instantons with
no moduli associated to positions in the four-dimensional space , i.e. no
$\Gamma$-invariant massless excitation of a string starting
and ending on the same D(-1)-brane.
Fractional instantons, unlike the regular ones,
have no complete images under the action of $\Z_p$  and therefore they cannot move away from the singularity.

We will start by considering the $U(1)$ case.
 The two main instanton classes are defined by the conditions
 \beqa
 {\rm Regular}:\quad && k_0=k_1=\ldots=k_{p-1}  \quad\quad r=0 \nn\\
 {\rm Fractional}:\quad && {\rm dim}_{\C} {\cal M}^{U(1)}_Y=0
  \label{fracreg}
\eeqa
For the sake of simplification in this $U(1)$ case we drop the subscript for the integer $r_\alpha$.
 It is important to notice that according to (\ref{ut}) regular instantons satisfy
 $u_0=N$, $u_1=u_2=...=0$, i.e. regular instantons carry always zero first Chern class.
 Instantons on $A_{p-1}$ can be built out of regular and fractional ones
\cite{Fucito:2004ry,Fujii:2005dk}. More precisely the instanton partition function
factorizes into a product of a contribution coming
from regular and fractional instantons
\beq
 Z_{A_{p-1}}=Z_{\rm frac} \, Z_{\rm reg}
\eeq
 This can be seen by noticing that the $\C^2$ partition function
can be rewritten as
\beqa
Z_{\C^2/\Z_2} &=& \bullet_0+\tinyyoung{0}
+\tinyyoung{1,0}+\tinyyoung{01}
+\tinyyoung{0,1,0}
+\tinyyoung{1,01}
+\tinyyoung{010}
+\tinyyoung{0,1,0,1}+\tinyyoung{0,1,01}
+\tinyyoung{10,01}
+\tinyyoung{1,010}+\tinyyoung{0101}
+\ldots\nn \\
&=& \left(\bullet_0
+\tinyyoung{0}
 +\tinyyoung{1,01}+\tinyyoung{0,10,010}+\ldots\right)
\left(\bullet_0+\tinyyoung{1,0}+\tinyyoung{01}
+\tinyyoung{0,1,0,1}+\tinyyoung{0,1,01}
+\tinyyoung{10,01}
+\tinyyoung{1,010}+\tinyyoung{0101}
+\ldots \right)\nn\\
&=& \left(1+q_0+q_0 \, q_1^2+q_0 \, q_1^2+q_0^4 \, q_1^2+\ldots \right)
\left(1+2\, q_0\, q_1+5\,q_0^2 \, q_1^2+\ldots \right)\label{zfp2}
\eeqa
or
\beqa
Z_{\C^2/\Z_3} &=& \bullet_0+\tinyyoung{0}
+\tinyyoung{2,0}+\tinyyoung{01}
+\tinyyoung{1,2,0}
+\tinyyoung{2,01}
+\tinyyoung{012}
+\tinyyoung{0,2,1,0}+\tinyyoung{2,1,01}
+\tinyyoung{20,01}
+\tinyyoung{2,012}+\tinyyoung{0120}
+\ldots \label{zfp3}\\
&=& \left(
\bullet_0+\tinyyoung{0}+\tinyyoung{01}+\tinyyoung{2,0}+
+\tinyyoung{2,012}+\tinyyoung{1,2,01}
+\tinyyoung{1,2,012}+\ldots\right)
\left(\bullet_0+\tinyyoung{1,2,0}+\tinyyoung{2,01}
+\tinyyoung{012}+\ldots \right)\nn\\
&=& \left(1+q_0+q_0 \, q_1+q_0 \, q_2+q_0 \, q_1\, q_2^2+ q_0 \, q_1^2\, q_2+q_0 \, q_1^2\, q_2^2+     \ldots \right)
\left(1+3\, q_0\, q_1\, q_2 +\ldots \right)\nn
\eeqa
and so on.
 The numbers in the boxes refer to the $\Z_p$ representation under which
the corresponding box (or D(-1)-instanton) transforms. A term $q_0^{k_0}\, q_1^{k_1}\,...$ represents
a harmonic form in the moduli space component with $k_a$ of $a$-type.
Here we take $r=0$ i.e. $N_a=\delta_{a,0}$.
The remaining
choices $r=1,2,\ldots,p-1$ can be found from this by performing a
cyclic permutations of the numbers in the starting boxes in $Y$ and $Y_{\rm frac}$.
Notice that a regular diagram starts always with a box ``0'' according to its definition (\ref{fracreg})
and its $k_a$'s are invariant under these cyclic permutations.
As shown in \cite{Fujii:2005dk}, the
pair $(Y_{\rm frac},Y_{\rm reg})$ can be alternatively  specified by the $p$ sets $(u_a,Y_a)$
with Young tableaux $Y_a$ and integers $u_a$ satisfying $\sum_{a=0}^{p-1}\, u_a=N$.
The regular and fractional part of a diagram $Y$ can be extracted following
a combinatoric algorithm developed in \cite{Fujii:2005dk}. This algorithm is reviewed
in appendix \ref{fuji}.

$U(N)$ instantons are built by tensoring $N$ copies of the $U(1)$-instanton
bundles. Now we start from the $N$ tableaux set $\{ (Y_\alpha,r_\alpha) \}$
in $\C^2$, compute the character and restrict to $\Z_p$-invariant
monomials. The tableaux $(Y_\alpha,r_\alpha)$ can be again decomposed into its regular
and fractional part  $(Y^{\rm reg}_{\alpha}|Y^{\rm
frac}_{\alpha},r_\alpha)$ with $Y^{\rm reg}_{\alpha}$, $Y^{\rm
frac}_{\alpha}$ tableaux of the regular and fractional type
respectively and $r_\alpha$ are integers ${\rm mod} p$.
Alternatively the same information can be encoded in the $p N$ set $(Y_{a\alpha}^*,u_{a\alpha})$
(see Appendix \ref{fuji} for details).

We say that a set $Y_\alpha$ is regular or fractional if all of the $Y_\alpha$'s are of the same type.
We remark that according to this definition, unlike the $U(1)$ case, the moduli space of a fractional
instanton has
not necessarily dimension zero since even for fractional tableaux $\Gamma$-invariant
terms can  appears in (\ref{charinv}) from terms with $Y_{{\rm frac},\alpha} \neq Y_{{\rm frac},\beta}$.
However these extra moduli are not associated to positions in $M_4$ (open string starting and ending on the same D(-1)-brane) and therefore fractional instantons are
stuck at the singularity as expected.

Using (\ref{chern2}), the instanton partition function (\ref{partfunct}) can be written as
\beqa
 {\cal Z}(q) &=& \sum_Y q^{k_0+\ft12 C^{aa} N_a}e^{-z^a\,u_a}
\eeqa
It is important to notice that the Chern characters
can be written as a sum over $U(1)$ contributions from $Y_\alpha$
\beqa
{\rm ch}_1({\cal E}_Y) &=&\sum_\alpha {\rm ch}_1({\cal E}_{Y_\alpha})
= \sum_{\alpha,a}  u_{a \alpha} {\rm ch}_1({\cal T}^a)\nn\\
k &=&k_0+\ft12 C^{aa} N_a
= \sum_\alpha \left(k_{0,\alpha}+\ft12 C^{r_\alpha r_\alpha} \right)\label{chernUN}
\eeqa
with $k_{a\alpha}$ the number of instantons in $Y_\alpha$
transforming in representation $R_a$ and
\beq
u_{a\alpha} = \delta_{r_\alpha,a}+k_{a+1,\alpha}+k_{a-1,\alpha}-2 k_{a,\alpha}
\label{ut2}
\eeq
Notice that $u_{a\alpha}$ satisfy the constraint
\beq
\sum_a u_{a\alpha}=1
\eeq
i.e. $u_{a\alpha}$ is characterized by $(p-1)N$ independent components $u_{a>0,\alpha}$.
According to the factorization algorithm explained before the instanton fixed points can be
completely characterized by the N-sets $\{ Y^*_{a\alpha},u_{a>0,\alpha} \}$ with
$Y^*_{a\alpha}$, $Np$ Young tableaux and a point in $u_{a\alpha}\in\Z^{N(p-1)}$.

 The decomposition (\ref{chernUN}) implies that the $U(N)$ partition function factorizes into
\beq
{\cal Z}_{U(N)}={\cal Z}_{U(1)}^N=({\cal Z}_{U(1),{\rm reg}}
{\cal Z}_{U(1),{\rm frac}} )^N
\eeq
 This is not surprising since the Euler number of the instanton moduli
space cannot depend on continuous deformations and therefore it
can be computed in the completely broken $U(1)^N$ phase when
D3-branes are far away from each other.

In the following we analyze the two type of instantons separately.

\subsection{Regular Instantons}

The moduli space of regular instantons can be described as follows.
First consider the character (\ref{chardef}) computed over the
ring of holomorphic polynomials $\C[z_1,z_2]$ on $\C^2$.
\beqa
\chi_\epsilon
(\C[z_1,z_2] ) ={1\over (1-T_1)(1-T_2)}=1+T_1+T_2+T_1^2+T_2^2+T_1 T_2+...
\eeqa
 Now we project the character onto those polynomials in $\C^2$ that
are invariant under $\Z_p$. The result can be rewritten in the
form
\beqa
\chi_\epsilon(\C^\Gamma[z_1,z_2] ) &=& {1\over p} \sum_{a=0}^{p-1}
{1\over (1-\omega^a\, T_1)(1-\omega^{-a}\, T_2)}\quad\quad
\omega=e^{2\pi i\over p}\nn\\
&=& \sum_{a=0}^{p-1} \, {1\over
(1-T_1^{p-a} T_2^{-a}) (1-  T_1^{1-p+a}\,T_2^{1+a})}\label{chiralap}
\eeqa
 This implies that the space $\C^2/ \Z_p$ can be thought of
as $p$ copies of $\C^2$ with coordinates $(z_{1a},z_{2a})$ in each chart
transforming as $( z_1^{p-a} z_2^{-a},   z_1^{1-p+a}\,z_2^{1+a})$.
 The character of the  instanton moduli space can therefore be written as a sum
over $p$ copies of the $\C^2$ character (\ref{trace})
\beq
 \chi_\epsilon({\cal M}_{Y_{\rm reg}})
= \sum_{a=0}^{p-1}\, \sum_{\alpha, \beta}^N\,  {\cal N}_{\alpha,\beta}^{Y^*_a}
(T_1^{p-a} T_2^{-a},   T_1^{1-p+a}\,T_2^{1+a})
\label{chinAp}
\eeq
 This implies in particular that regular instantons are specified
 by $p$ N Young tableaux $Y_a^*=\{ Y^*_{a\, \alpha} \}$.
 The $Y_{a}^*$ follows from $Y_{\rm reg}$ via the factorization algorithm explained in the last section.
The instanton partition function computed using either
 (\ref{chinAp}) or (\ref{charinv}) leads to the
same results but expression (\ref{chinAp}) can be easily resummed
to all instanton orders even in  the general
$\C^2/\Gamma_{p,q}$ case. 
In fact counting fixed points in the regular instanton moduli space character
(\ref{chinAp})  boils down to count the number of partitions, $Y^*_{a\alpha}$, of the integers $k_{a\alpha}$.
The precise relation between the two
descriptions will be explained in Appendix \ref{fuji}.

Let us now compute the regular instanton partition function.
First notice that $k_0=k_1=..=k_{p-1}=k$ implies $u_a=0$, i.e. regular instantons
have zero first Chern character and instanton number
\beq
 k ({\cal M}_{\rm reg})=\sum_{a,\alpha} |Y^*_{a \alpha}|
\eeq
where by $|Y^*_{a \alpha}|$, we denote the number of boxes in $Y^*_{a \alpha}$.
 The partition function is then given by the $Np$ power of the number of Young tableaux, i.e.
 the number of partitions of $k$
 \beq
 Z_{\rm reg}={1\over \hat{\eta}(q)^{N p}}  \quad\quad
 \hat{\eta}(q)\equiv \prod_{n=1}^\infty (1-q^n)\label{regzp}
 \eeq
Actually, we can do better and consider the cohomology of these spaces. For simplicity we take $N=1$.
 As we explained
before the Betti numbers $b_{2 n_-}={\rm dim} H^{2 n_-}({\cal M}_{\rm reg})$
are given in terms of the number $n_-$ of negative eigenvalues
in (\ref{chinAp}) with the conditions (\ref{range}).
 A simple inspection of (\ref{trace}) shows that negative eigenvalues
can come only from the term
 \beq
    T_1^{-(p-a)h(s)-
(v(s)+1)(p-a-1)}\, T_2^{a h(s)+
 (v(s)+1)(1+a)}
\eeq
 The eigenvalue happens to be negative whenever any of
the following conditions is satisfied
\begin{itemize}
\item{ $a\neq p-1$}
\item{$a=p-1~{\rm and}~h(s)>0$}
\end{itemize}
A row of length $m$ in
the $Y_\alpha$ tableau contributes then to the Poincar\'{e} Polynomial
$q^m\,t^{2N m}$ when $a\neq p-1$ and  $q^m\, t^{2N(m-1)}$
when $a=p-1$. The generating function for the
Poincar\'{e} polynomial can then be written as
\beqa
 P(t|{\cal M}^{U(1)}_{\rm reg})
&=& \prod_{m=1}^\infty {1\over (1-q^m\, t^{2 m})^{p-1} (1-q^m\,
t^{2 m-2})}= \sum_{n=0}^\infty q^n\,P(t|(\C^2/\Z_p)^n/S_n )\nn
\eeqa
 in agreement with the fact that the moduli space of
regular $U(1)$ instantons on $\C^2/\Z_p$ is given by the Hilbert
scheme of points on $\C^2/\Z_p$ with $b_0=1,b_2=p-1$ (see \cite{nakajima}
for the Poincar\'{e} polynomial symmetric product formula).

\subsection{Fractional Instantons }

Now we consider fractional instantons. We start with the $U(1)$ case.
Fractional instantons correspond to Young tableaux containing no
$Z_p$-invariant box i.e. those tableau with no box $s=(i,j)$ with hook length
satisfying $\ell(s)=\nu_{j}+ \tilde{\nu}_i-i-j+1=0 ~{\rm mod}~ p$.
Alternatively fractional instantons are defined by the condition
\beqa {\rm dim}_{\C} {\cal M}^{U(1)}_Y &=& 0 \quad\quad \Rightarrow \quad\quad
k_{r} = \ft12\widehat{C}_{a b}\, k_a\, k_{b} \label{dim0}
\eeqa
 This condition can be used to rewrite the instanton number $k$ for fractional
instantons entirely in terms of their first Chern classes $u_a$. To see
this we first relate the dimension of the instanton moduli space
to the square of its first Chern class \footnote{Here we used the
identities
$C^{ab}\widehat{C}_{a0}=C^{ab}\widehat{C}_{a0}\widehat{C}_{b0}=1$.}
\beqa
C^{a b}\, u_a\, u_{b} &=& C^{ab} N_a N_b+C^{ab} \widehat{C}_{bc} \widehat{C}_{bd} k_c k_d-
2 C^{ab} \widehat{C}_{bc} k_c N_d\nn\\
&=& C^{ab} N_a N_b+\widehat{C}_{ab} k_a k_b-2 k_a N_a+2 k_0 N\nn\\
&=& C^{ab} N_a N_b +2 k_0 N -{\rm dim}_{\C} {\cal M}_Y \label{relation}
\eeqa
 Then specializing to $U(1)$ fractional instantons, i.e. $N=1$ and ${\rm dim}_{\C} {\cal M}_Y=0$,
one finds
\beq
k=k_0+C^{aa} N_a=\ft12\,C^{a b}\, u_a\, u_{b}\label{kuu}
\eeq
 We recall that  $u_a$ spans $\Z^{p-1}$.

It is instructive to work it out some explicit examples.
For $p=2$ one finds
 \beqa
Z_{\rm frac,N=(1,0)}^{\Z_2} &=&\bullet_0
+\tinyyoung{0}
 +\tinyyoung{1,01}+\tinyyoung{0,10,010}+\tinyyoung{1,01,101,0101}
+\ldots\nn\\
%&=& 1+ q_0 +q_0 \,q_1^2 + q_0^4\, q_1^2 + q_0^4\, q_1^6\ldots\nn\\
Z_{\rm frac,N=(0,1)}^{\Z_2} &=&\bullet_1
+\tinyyoung{1}
 +\tinyyoung{0,10}+\tinyyoung{1,01,101}+\tinyyoung{0,10,010,1010}
+\ldots
%\nn\\
%&=& 1+ q_1 +q_1 \,q_0^2 + q_1^4\, q_0^2 + q_1^4\, q_0^6\ldots
\label{Yf}
\eeqa
 Evaluating $u_1$
\beq
u_1=2(k_0-k_1)+N_1
\eeq
  for the diagrams in (\ref{Yf}) one finds
\beqa
r=0 && \quad\quad u_1=0,2,-2,4,-4,\ldots\nn\\
r=1 && \quad\quad u_1=1,-1,3,-3,5,\ldots
 \eeqa
 i.e. $u_1$ spans $\Z$ in agreement with our general claim.
The fractional instanton partition function can then be written as
\beq
{\cal Z}_{\rm frac}^{\Z_2}=\sum_{u\in \Z} q^{\ft14 u^2}\, e^{-z^a u_a}
\eeq
 For $p=3$, $r=0$ one finds
\beqa
Z_{\rm frac,N=(1,0,0)}^{\Z_3}
&=&\bullet_0+\tinyyoung{0}+\left(\tinyyoung{01}+\tinyyoung{2,0}\right)
+\left(\tinyyoung{2,012}+\tinyyoung{1,2,01}\right)
+\tinyyoung{1,2,012}+\ldots
%\nn\\
%&=& 1+q_0+q_0 q_1+ q_0 q_2+ q_0 q_1 q_2^2+q_0 q_1^2 q_2+q_0 q_1^2 q_2^2
%+\ldots
\label{z3}
\eeqa
 The results for $r=1,2$ follows from (\ref{z3}) by
cyclic permutations of the labels. One can easily check that
 the resulting spectrum of $u_a=N_a+k_{a+1}+k_{a-1}-2 k_a$'s spans $\Z^2$.
 %r=0 &:&(0,0),(1,1),(-1,2),(2,-1),(1,-2),(-2,1),(-1,-1),\ldots\nn\\
%r=1 &:&(1,0),(-1,1),(0,-1),(0,2),(2,1),(2,-2),(3,-1),\ldots\nn\\
%r=2 &:&(0,1),(1,-1),(2,0),(-1,0),(-2,2),(1,2),(1,-3),\ldots
%\eeqa

 Using (\ref{kuu}) the instanton partition function can then be written in general as
\beq
{\cal Z}_{\rm frac}(q)=
\sum_{u\in \Z^{p-1}} q^{ \ft12 C^{ab} u_a u_b} \, e^{-z^a u_a}\label{resu1}
\eeq

The result for $U(N)$ is then given by the $N$ power of (\ref{resu1})
\beq
{\cal Z}_{U(N), {\rm frac}}=
\sum_{ \vec{u}_{a}\in \Z^{N(p-1)}} q^{ \ft12 C^{ab} \vec{u}_{a}\cdot
\vec{u}_{b}}\, e^{-z^a u_a} \label{resap}
\eeq
 with $  \vec{u}_{a}\cdot
\vec{u}_{b}\equiv  \sum_\alpha  u_{a\alpha}\,
u_{b\, \alpha }$ and $u_a=\sum_\alpha u_{a\alpha}$.
This result has been anticipated in \cite{Fujii:2005dk} (see also  \cite{Bianchi:1996zj}
for the $U(1)$ case).

\section{Instantons on toric varieties}
\label{sop}
\setcounter{equation}{0}

Here we consider instantons on more general toric
varieties. We start from the singular case.
The most general toric singularity in four dimensions can
be written as the quotient $\C^2/\Gamma_{p,q}$,
with $\Gamma_{p,q}$ a $\Z_p$ action specified by the two coprime integers $p,q<p$ with generator
\beq
\Gamma_{p,q}:~~~~  \left(%
\begin{array}{c}
  z_1 \\
  z_2 \\
\end{array}%
\right) \rightarrow  \left(%
\begin{array}{cc}
  e^{2\pi i /p} & 0 \\
  0 &   e^{2\pi i q/p}\\
\end{array}%
\right)\left(%
\begin{array}{c}
  z_1 \\
  z_2 \\
\end{array}%
\right)
\eeq
 and $z_{1,2}$ the complex coordinates on $\C^2$.
The two extreme cases $q=p-1$ and $q=1$ correspond to
the familiar $A_{p-1}$ and blown down ${\cal O}_{\PP_1}(-p)$ surfaces respectively.

For generic $(p,q)$ a smooth manifold is obtained by blowing up points to exceptional
surfaces , $C_i$, whose intersection numbers are given expanding the fraction $p/q$ as
\beq
{p\over q}=e_1-{1\over e_2-{1\over \ldots-{1\over e_n}}}\label{junk}
\eeq
 in terms of the integers $e_i$. To each $i=1,2,\ldots,n$ one associates the
two-cycle $C_i$ with self-intersection $e_i$.
 The intersection matrix is then given by \cite{barth}
 \beq
C= \left(%
\begin{array}{cccccc}
  e_1 & -1 & 0 & \ldots& \ldots & 0 \\
  -1 & e_2 & -1 & 0&\ldots & 0\\
  0 & -1 & e_3 & -1 & \ldots& \ldots \\
\ldots & \ldots & \ldots & \ldots & \ldots& \ldots \\
  0 & \ldots & \ldots& 0 & -1 & e_n \\
\end{array}%
\right)\label{cint}
 \eeq
The two extreme cases $q=p-1$ and $q=1$ lead to $n=p-1,e_{i}=2$ and $n=1,e_1=p$ respectively
justifying the identification with the $A_{p-1}$ and
the blown down ${\cal O}_{\PP_1}(-p)$ surfaces. For a general $(p,q)$ one
finds a ``necklace'' of $n$ two-spheres.
 This space can be covered by $n+1$ charts each looking locally like $\C^2$.

 We will consider $U(N)$ instantons on these spaces. Unlike the $A_{p-1}$
case a description of the ADHM instanton moduli space for
a general $\Gamma_{p,q}$-singularity with $q\neq p-1$
is not available in the literature
(see \cite{Nakajima:2003pg,Nakajima:2003uh,Sasaki:2006vq} for results in the $O_{\PP_1}(-p)$ case for
$p=1,2$).
 Still regular and fractional instantons
on $M=\C^2/ \Gamma_{p,q}$ can be easily constructed.
  As before, the $U(N)$ instanton partition function can be written in terms
  of that of $U(1)$ and therefore we can restrict ourselves to the $U(1)$ case.
Regular $U(1)$ instantons are instantons transforming in the regular
representation of $\Z_p$. They can move freely on $M$ in sets of
$p$-images symmetrically distributed around the singularity.
The moduli space of k regular instantons is then given by specifying k-points
on $M$ up to permutations i.e. the Hilbert scheme $M^{[k]}$.
Their contribution to the partition function is given by
\beq
Z^{U(1)}_{\rm reg}=\sum_k\, q^k \chi(M^{[k]})\equiv\hat{\eta}(q)^{-\chi(M)} \quad\quad
\hat{\eta}(q)=\prod_{n=1}^\infty (1-q^n)
\eeq
 Next we consider instantons carrying non-trivial first Chern-class
 along the exceptional two-cycles.
  A self-dual $U(1)$ gauge field strength can be written as
 \beq
{ {\rm i}F\over 2\pi}=u_i \,\alpha^i  \in H_2^+(M,\Z)
\eeq
 with $\alpha^i$, $i=1,..b_2^+$ a basis $H_2^+(M,\Z)$ and $u_i$ some integers
 \footnote{In the $A_{p-1}$ case one can choose $\alpha^i=c_1({\cal T}^i)$
 as the basis.}. These gauge
 connections are self-dual by construction and corresponds to
 isolated points in the  moduli space since they do not
 admit any continuous deformation. We call them "fractional".

Their contribution to the Yang Mill action with the insertion of
the observables in  (\ref{partfunct}) can be written as
 \beqa
 S_{SYM} &=& -{i \,\tau\over 4\pi }\, \int_M  F\wedge F
 -\varphi_{2i} \, \int \,  { {\rm i}F\over 2\pi}\wedge \alpha^i\nn\\
 &=& -\pi\, i\, \tau\, C^{ij} u_i u_j+z^i\, u_i
 \eeqa
 with $\tau={4\pi i\over g_{\rm YM}^2}+{\theta\over 2 \pi}$ and
$z^i=C^{ij}\, \varphi_{2j}$. The fractional instanton partition
function $\sum_{u} e^{-S_{YM}}$ can then be written as
\beq
Z^{U(1)}_{\rm frac}
 =\sum_{u_{i}\in \Z^{b_2^+(M)}} q^{ {1\over 2} C^{ij} u_{i}
\, u_{j}}\, e^{- z^i\,u_i }
\eeq
Collecting the regular and fractional instanton contributions
 one finds the partition function
 \beq
 Z=(Z^{U(1)}_{\rm reg} \, Z^{U(1)}_{\rm frac})^N
 ={1\over \hat{\eta}(q)^{N\, \chi(M)}}   \,
\sum_{ \vec{u}_{i}\in \Z^{N\, b_2^+(M)}} q^{ {1\over 2} C^{ij} u_{i\alpha}
\, u_{j\alpha}}\, e^{- z^i\,u_i } \label{zbh}
 \eeq
 where $u_i=\sum_{\alpha=1}^N u_{i\alpha}$,
$\chi(M)$, $b_2^+(M)$ are the Euler number and the number of self-dual
 forms in $M$ respectively and $C^{ij}$ is the inverse of the intersection
 matrix given by (\ref{cint}).
 Specifying to the $A_{p-1}$ case,
$\chi=b_2^+ +1=p$ one finds the result (\ref{resu1}).

In this paper we mainly focus on non-compact toric varieties but we should
stress that our considerations extend naturally to the compact case.
 Indeed, in general a toric variety can be covered by $\chi(M)$ charts each looking
 locally as $\R^4$ and one can think of regular $U(N)$ instantons on $M$ as
 $\chi(M)$ copies of $U(N)$ instantons on $\R^4$. Equivalently they can be
 thought of as $N$ copies of $U(1)$ instantons described by the Hilbert schemes $M^{[k]}$.
 Their contribution to the partition formula is $\hat{\eta}^{-\chi N}$.
Fractional instantons can be built out of integer valued
linear combinations  $F=u_i \,\alpha^i$ of the self-dual forms $\alpha^i$ forming
a basis $H^+_2(M,\Z)$ and contribute to the lattice sum in (\ref{zbh}).

For a general toric variety the ADHM construction of self-dual connections
is not known. In general we cannot ensure that any self-dual connections
can be entirely built out of regular and fractional instantons of the type
describe here and therefore extra contributions to 
(\ref{zbh}) cannot be excluded.
Nonetheless the tests we will perform against 2d SYM computations
 and supergravity D0-D2-D4 black entropy formulae suggest that (\ref{zbh}) holds
 (even for compact manifolds) at least in the limit where instanton
 charges are taken to be large.

\subsection{Instantons on $O_{\PP_1}(-p)$}

As an illustration consider the case $q=1$: the blown
down $O_{\PP_1}(-p)$ surface.

{\bf Regular instantons}

The ring of invariant polynomials in $\C^2$ under
$\Gamma=\Gamma_{p,1}$ can be written as \beqa \chi_\epsilon
(\C^\Gamma[z_1,z_2] ) &=& {1\over p} \sum_{a=0}^{p-1} {1\over
(1-\omega^a\, T_1)(1-\omega^{a}\, T_2)}\quad\quad
\omega=e^{2\pi i\over p}\nn\\
&=& {1\over
(1-T_1^{p}) (1-{T_2\over  T_1 })}+{1\over
(1-T_2^{p}) (1- {T_1\over  T_2 })}\label{chiralop}
\eeqa
 This implies that the space $\C^2/\Gamma$ can be thought of
as two copies of $\C^2$ with coordinates
$( z_1^{p}, {z_2\over z_1})$ and $( {z_1\over z_2},z_2^p)$ in the two patches.
 The instanton moduli space character can therefore be written as a sum
of two $\C^2$ characters
\beq
 \chi_\epsilon({\cal M}_{Y_{\rm reg}})
=  \sum_{\alpha, \beta}^N\,\left[ {\cal N}^{Y_1}_{\alpha,\beta}
\left(T_1^{p}, {T_2\over  T_1 } \right)+{\cal N}^{Y_2}_{\alpha,\beta}
\left( {T_1\over T_2},T_2^p \right) \right]
\label{chinOp}
\eeq
 specified by the $2N$ Young tableaux
$\{ Y_{1\alpha},Y_{2\alpha}\}$.
The partition function is then given by the $2N$ power of the number of partitions of $k$
 \beq
 Z_{\rm reg}={1\over \hat{\eta}(q)^{2 N}}   \label{regop}
 \eeq
The Poincar\'{e} polynomial can be computed as before counting the number
of negative eigenvalues in (\ref{chinOp}).
 For $N=1$ there is a negative eigenvalue whenever any of
the following conditions are satisfied
\begin{itemize}
\item{$Y_1$ : for each box }
\item{$Y_2$: $h(s)>0$}
\end{itemize}
  The Poincar\'{e} generating polynomial can then be written as
\beqa
 P(t|{\cal M}^{U(1)}_{\rm reg})
&=& \prod_{m=1}^\infty {1\over (1-q^m\, t^{2 m}) (1-q^m\,
t^{2 m-2})}= \sum_{n=0}^\infty q^n\,P(t|(\C^2/\Gamma)^n/S_n )\nn
\eeqa
 in agreement with the fact that the moduli space of
regular $U(1)$ instantons on $\C^2/\Gamma_{p,1}$ is given by the Hilbert
scheme of points on $\C^2/\Gamma_{p,1}$ with $b_0=b_2=1$.

{\bf Fractional instantons}

Fractional instanton connections can be constructed as before in terms
of integer-valued combinations of the self-dual two forms on $\C^2/\Gamma$.
In the $O_{\PP_1}(-p)$ there is a single two-form $c_1({\cal T})$ with
self-intersection
\beq
\int c_1({\cal T})\wedge c_1({\cal T})=-{1\over p}
\eeq
 The fractional self dual connection can then be written as
\beq
{{\rm i}F \over 2 \pi}={\rm diag}\, (u_1,u_2,...u_N) \, c_1({\cal T})
\eeq
 with Yang-Mill action
\beqa
 S_{SYM} &=& -{i\, \tau\over 4\pi }\, \int_M  {\rm tr}\,F\wedge F
 + \varphi_{2}\, \int {\rm tr}\, \, {{\rm i} F\over 2\pi}\wedge c_1({\cal T})\nn\\
 &=& -{\pi\, i\, \tau\over p}\, \vec{u} \cdot\vec{u}
 +{1\over p}\, u \, \varphi_{2}
 \eeqa
 and $u=\sum_\alpha u_\alpha$. The partition function can then be written as
\beq
{\cal Z}_{\rm frac}=
\sum_{u_\alpha\in \Z^{N}} q^{ \frac{1}{2p} \vec{u}\cdot \vec{u}}
e^{-z u}\quad\quad z={\varphi_2\over p} \label{resu3}
\eeq

\subsection{Instantons on $\C^2/\Gamma_{5,2}$}

Finally we consider the singularity $\C^2/\Gamma_{5,2}$.
This singularity does not belong neither to the $A_{p-1}$ nor to the $O_{\PP_1}(-p)$
series and illustrates the general case.

{\bf Regular instantons}

The ring of invariant polynomials in $\C^2$ under
$\Gamma=\Gamma_{5,2}$ can be written as \beqa \chi_\epsilon
(\C^\Gamma[z_1,z_2] ) &=& {1\over p} \sum_{a=0}^{4} {1\over
(1-\omega^a\, T_1)(1-\omega^{2 a}\, T_2)}\quad\quad
\omega=e^{2\pi i\over 5}\nn\\
&=&
{1+T_1 T_2^2+T_1^2 T_2^4+T_1^3 T_2+T_1^4 T_2^3\over
(1-T_1^5) (1-T_2^5)}\nn\\
&=& {1\over (1-T_1^5) (1-{T_2\over  T_1^2 })}
+{1\over (1-{T_1^2\over  T_2}) (1- {T_2^3\over  T_1 })}
+ {1\over (1-{T_1\over  T_2^3 })(1-T_2^{5})}\label{chiral52}
\eeqa
 This implies that the space $\C^2/\Gamma_{5,2}$ can be thought of
as three copies of $\C^2$ with coordinates
$(z_1^5,{z_2\over  z_1^2 })$, $\left({z_1^2\over  z_2},{z_2^3\over  z_1 } \right)$
and $\left( {z_1\over  z_2^3},z_2^{5} \right)$ on the three patches.

 The regular instanton moduli space character can therefore be written as a sum
of three $\C^2$ characters
\beq
 \chi_\epsilon({\cal M}_{Y_{\rm reg}})
=  \sum_{\alpha, \beta}^N\,\left[ {\cal N}^{Y_1}_{\alpha,\beta}
\left(T_1^5,{T_2\over  T_1^2 } \right)+{\cal N}^{Y_1}_{\alpha,\beta}
\left({T_1^2\over  T_2},{T_2^3\over  T_1 } \right)+{\cal N}^{Y_2}_{\alpha,\beta}
\left( {T_1\over  T_2^3},T_2^{5} \right) \right]
\label{chinOp1}
\eeq
 specified by the $3N$ Young tableaux
$\{ Y_{1\alpha},Y_{2\alpha},Y_{3\alpha}\}$.
The partition function is then given by the $3N$ power of the number of partitions of the integer $k$
 \beq
 Z_{\rm reg}={1\over \hat{\eta}(q)^{3 N}}   \label{regop1}
 \eeq
The Poincar\'{e} polynomial can be computed as before counting the number
of negative eigenvalues in (\ref{chinOp1}).
 For $N=1$ there is a negative eigenvalue whenever any of
the following conditions are satisfied
\begin{itemize}
\item{$Y_1,Y_2$ : for each box }
\item{$Y_3$: $h(s)>0$}
\end{itemize}
  The generating function of the Poincar\'{e}  polynomial can then be written as
\beqa
 P(t|{\cal M}^{U(1)}_{\rm reg})
&=& \prod_{m=1}^\infty {1\over (1-q^m\, t^{2 m})^2 (1-q^m\,
t^{2 m-2})}= \sum_{n=0}^\infty q^n\,P(t|(\C^2/\Gamma_{5,2})^n/S_n )\nn
\eeqa
 in agreement with the fact that the moduli space of
regular $U(1)$ instantons on $\C^2/\Gamma_{5,2}$ is given by the Hilbert
scheme of points on $\C^2/\Gamma_{5,2}$ with $b_0=1$, $b_2=2$.

{\bf Fractional instantons}

Fractional instanton connections can be constructed as before in terms
of integer-valued combinations of the self-dual two forms $c_1({\cal T}^i)$, $i=1,2$,
on $\C^2/\Gamma_{5,2}$ with self-intersection
\beq
\int c_1({\cal T}^i)\wedge c_1({\cal T}^j)=-C^{ij}
\eeq
 with $C^{ij}$ the inverse of (\ref{cint})
 \beq
 C_{ij}=\left(%
\begin{array}{cc}
  3 & -1 \\
  -1 & 2 \\
\end{array}%
\right)\quad\quad C^{ij}=\ft15\left(%
\begin{array}{cc}
  2 & 1 \\
  1 & 3 \\
\end{array}%
\right)
 \eeq

 The fractional self dual connection can then be written as
\beq
{{\rm i}F \over 2 \pi}={\rm diag}\, (u_{i1},u_{i2},...u_{iN}) \, c_1({\cal T}^i)
\eeq
 with Yang-Mill action
\beqa
 S_{SYM} &=& -{i\, \tau\over 4\pi }\, \int_M  {\rm tr}\,F\wedge F
 + \varphi_{2i}\, \int {\rm tr}\, \, {{\rm i} F\over 2\pi}\wedge c_1({\cal T}^i)\nn\\
 &=& -\pi\, i\, \tau\, C^{ij}\, \vec{u}_i \cdot\vec{u}_j
 +C^{ij}\, u_i \, \varphi_{2j}
 \eeqa
 and $u_i=\sum_\alpha u_{i\alpha}$. The partition function can then be written as
\beq
{\cal Z}_{\rm frac}=
\sum_{u_{i\alpha}\in \Z^{2N}} q^{ \frac{1}{2}\, C^{ij}\, \vec{u}_i\cdot \vec{u}_j}
e^{-z^i u_i}\quad\quad z^i=C^{ij}\, \varphi_{2j} \label{resu31}
\eeq

\section{Black hole partition functions}
\label{sbh}
\setcounter{equation}{0}

In this section we derive a microscopic formula for the
partition function of a black hole made out of
D4-D2-D0 bound states wrapping a four cycle inside a
CY. We will restrict ourselves to the case where both
the cycle and the CY are compact.
The lift of this brane system to M-theory is well known and
a microscopic derivation of the corresponding black hole entropy
based on a two-dimensional $(4,0)$ SCFT has been derived in
\cite{msw} (see also
\cite{Vafa:1997gr,LopesCardoso:1998wt,LopesCardoso:1999ur,LopesCardoso:1999xn,Gaiotto:2005rp}
for related (macro)microscopic countings).
The aim of this section is to test our instanton
partition function formula against supergravity. We refer the readers to \cite{msw} for
details on the geometrical tools we will use here. We consider a single D4-brane wrapping
a very ample divisor P inside a CY.
The conjugacy class $[P]\in H^2(CY,\Z)$ can be expanded as $[P]=p^A \alpha_A$
with $\alpha_A$ a basis in $H^2(CY,\Z)$.

According to \cite{Ooguri:2004zv} the black hole partition function is defined
as
\beq Z_{BH}=\sum_{Q_0,Q_{A}} \, \Omega(Q_0,Q_{A},p^A) \,
e^{-Q_0 \varphi_0-Q_{A}\varphi^{A}}
\eeq
 with $\Omega(Q_0,Q_{A},p^A)$ the multiplicity of a bound state of $Q_0$
 D0-branes, $Q_{A}$ D2-branes and a D4 brane wrapping $P=p^A \Sigma_A$.
 $\varphi_0, \varphi^{A}$ are the D0,D2 chemical potentials.
 D0,D2 branes can be thought of as instantons and fluxes respectively
 in the worldvolume theory of the D4-brane
 \beqa
 && Q_0=k={1\over 8 \pi^2}\, \int_M {\rm tr}\, F\wedge F\quad\quad
 Q_{A}=C_{AB} u^B={1\over 2\pi}\,\int \, {\rm tr}\,F\wedge \alpha_A
 \eeqa
 Self-duality implies that $Q_{A>b_2^+}=0$.

  The black hole partition function can be then read from the instanton partition
  function formula (\ref{zbh})
  \beqa
  Z_{BH}
  %&=&{\cal Z}(P|q,z_A)=\sum_{k,u^A}\, \chi({\cal M}_{k,u^A}) \, q^k \, e^{-z_A\, u^A}
  %\nn\\
  & =& {1\over \hat{\eta}(q)^{\chi(P)}}
 \sum_{ u^A\in \Z^{b_2^+(P)}} q^{ {1\over 2} C_{AB} \,u^A\, u^B} e^{- \varphi_A\,u^A }
 = {1\over \hat{\eta}(e^{-\varphi_0})^{\chi(P)}}
 \sum_{ Q_A\in \Z^{b_2^+(P)}} e^{ -{1\over 12} D^{AB} \,Q_A\, Q_B\, \varphi_0- \varphi^A\,Q_A }
 \nn\\
&=& \sum_{Q_0,Q_{A}} \, \Omega(Q_0,Q_{A},p^A) \, q^{Q_0}\, e^{-\varphi^A Q_A}
 \label{finalzbh}
  \eeqa
with
\beqa
\chi(P) &=& \int_P\, c_2(P)=D_{ABC} p^A p^B p^C+c_{2A} p^A\nn\\
b_2^+(P) &=& 2\,D_{ABC} p^A p^B p^C+\ft16 c_{2A} p^A\nn\\
D_{ABC} &\equiv& \ft16 \int_{CY} \alpha_A\wedge \alpha_B\wedge \alpha_C \quad\quad
 c_{2A}\equiv\int_{CY} \alpha_A\wedge c_2(CY)\nn\\
C_{AB} &=&-\int_{P} \alpha_A\wedge \alpha_B=-6\,D_{AB}\quad\quad
D_{AB} \equiv D_{ABC}\, p^C\nn\\
q &=& e^{-\varphi_0} \quad\quad e^{-z_A}=e^{-C_{AB}\varphi^B}
\eeqa
and $C^{AB},D^{AB}$ the inverse of $C_{AB},D_{AB}$ respectively.

 Notice that (\ref{finalzbh}) is the partition function of $\chi(P)$ free bosons ($b_2^+$ of them 
living in the lattice $H^2(P,\Z)$) in two-dimensions.
 The black hole entropy follows from the Cardy formula
 \beqa
S_{BH} &\approx& {\rm ln}\, \Omega(Q_0,Q_{A},p^A)
\approx 2\pi \sqrt{\ft16\chi(P)Q_{0, {\rm reg}} }\nn\\
&=& 2\pi \sqrt{(D_{ABC} p^A p^B p^C+\ft16\,c_{2A} p^A )(Q_0+\ft{1}{12} D^{AB} Q_A Q_B)}
\label{entd0d4}
\eeqa
 with $Q_{0, {\rm reg}}=Q_0+\ft{1}{12} D^{AB} Q_A Q_B$ the number of regular instantons coming
 from the expansion of $\hat{\eta}^{-\chi}$ in (\ref{finalzbh}).
(\ref{entd0d4}) agrees with the micro/macroscopic M5-brane/supergravity results found in \cite{msw}.

\section{Summary and conclusions}
\label{sconcl}
\setcounter{equation}{0}

In this paper we have studied instantons on toric varieties.
Starting from the ADHM construction of instanton on $A_{p-1}$
we derived the instanton partition function formula
\beq
 Z=(Z^{U(1)}_{\rm reg} \, Z^{U(1)}_{\rm frac})^N
 ={1\over \hat{\eta}(q)^{N\, \chi(M)}}   \,
\sum_{ \vec{u}_{i}\in \Z^{N\, b_2^+(M)}} q^{ {1\over 2} C^{ij} u_{i\alpha}
\, u_{j\alpha}}\, e^{- z^i\,u_i } \label{zbh2}
 \eeq
  where $u_i=\sum_{\alpha=1}^N u_{i\alpha}$, $\chi(M)=p$, $b_2^+(M)=p-1$ are
  the Euler number and the number of self-dual
 forms in $M$ respectively and $C^{ij}$ is the inverse of the intersection
 matrix i.e. minus the Cartan matrix of $A_{p-1}$.
That the Euler number of the moduli space of $U(N)$ instantons
can be related to that of $U(1)$ instantons follows from the fact that
that after localization $U(N)$ is broken to $U(1)^N$.
 There are two very special classes of $U(1)$ instantons: {\it regular} and {\it fractional}. 
Regular instantons are instantons transforming in the regular
representation of $\Z_p$. They can move freely on $M$
in sets of $p$-images symmetrically distributed around the singularity.
The moduli space can then be thought of as the Hilbert scheme of $k$ points
in $M$ contributing $\eta^{-\chi}$ to the partition function.
Alternatively, covering $M$ with $\chi(M)$ charts, $U(N)$ instantons on
$M$ can be thought as $\chi(M)$ copies of $U(N)$ instantons on $\R^4$ each
contributing $\hat{\eta}^{-N}$.

Fractional instantons are instantons with no or incomplete images.
They are stuck at the singularity since all open strings moduli starting and ending
on the same D(-1)-brane are projected out by the orbifold. They carry non-trivial
fluxes along the exceptional two cycles and correspond
to self-dual connections ${iF\over 2\pi}=u_a \alpha^a$ 
where the $\alpha^a$ form a basis in $H^{2+}(M|\Z)$.
Their contribution to the partition function gives the lattice sum in (\ref{zbh2}).

For a general toric variety $M$, an ADHM construction is missing but still
regular and fractional instantons can be constructed exactly in the same
way as in the $A_{p-1}$ case\footnote{The fact that the instanton
partition function factorizes into a product of regular and fractional
instantons has been experimentally observed in \cite{Fucito:2004ry} for
instantons $A_{p-1}$-singularity and  shown in \cite{Fujii:2005dk}.
For general $(p,q)$, there is no proof that this should
be the case and we don't exclude that extra contributions can arise from
new classes of instantons. The perfect agreement with 2d SYM computations
and supergravity however suggests that this is not the case. }.
Here we work out in details the case of toric singularities of type
$M=\C^2/\Gamma_{p,q}$ with $\Gamma_{p,q}$ a $\Z_p$ action specified by
the coprime integers $(p,q)$. The blown down $O_{\PP_1}(-p)$ surface
corresponds to the case q=1, $\chi(M)=b_2^++1=2$
 It is interesting to notice that the result (\ref{zbh2})
 corresponds to the chiral partition function of a two-dimensional conformal field theory
 with central charge $N\chi$. This is not surprising from the M-theory
 perspective where the D4-D2-D0 quantum mechanics lift to a $(4,0)$
 two-dimensional SCFT living on the M5-brane worldvolume \cite{msw}.

Comparing our results with previous computations based on 2d SYM
\cite{Aganagic:2004js,Aganagic:2005wn,Caporaso:2005np} one finds
that only the contributions of fractional instantons are captured
by the 2d analysis. This is not surprising since the 2d SYM
picture focalizes on the neighborhood of the singularity and can
hardly trace regular instantons that can move freely far away from
the singularity.
 For the $U(1)$ case, the fractional instanton lattice sum in
 (\ref{zbh2}) is in perfect agreement with the 2d SYM results
in the case of  $A_{p-1}$ and $O_{\PP_1}(-p)$ surfaces.
More precisely taking $\chi(M)=p$ and $\chi(M)=2$ in (\ref{zbh2}) one finds
the Poisson resummation version of formula(6.7) in \cite{Aganagic:2005wn} and
formula (27) of \cite{Caporaso:2005np} respectively.
 Formula (\ref{zbh2}) extends these results to the general $(p,q)$-case
  and completes them with the inclusion of regular instanton
 contributions that are crucial to reproduce the right black hole entropy.

 The comparison for $N>1$ is trickier. The formulae in
 \cite{Aganagic:2005wn,Caporaso:2005np} include perturbative corrections in the gauge coupling, $g_{YM}$,
which cannot be interpreted as a topological number such as the Euler number of an instanton moduli space.
These corrections come from the Chern-Simon contributions on the boundary of the non-compact
space $\C^2/\Gamma_{p,q}$. Their significance has been very recently clarified in  \cite{griguolo}.
The two-dimensional SYM partition function has been written in a product form involving
an instanton sum  and a Chern-Simon contribution  from the boundary
Lens space in \cite{Caporaso:2005np,griguolo}.
Remarkably, discarding the semiclassical fluctuations in the Chern-Simon theory, the two results
perfectly agree, see (3.23) of \cite{griguolo}.

Finally we compute the partition function of D0-D2-D4 bound state wrapping
a compact four cycle $M$ with conjugacy
class $[M]=p^A\alpha_A \in H^2(CY|\Z)$ inside a CY. Applying the
Cardy formula to estimate  the growing of the bound state multiplicities in (\ref{zbh2}),
at large $Q_0$ one finds
\beqa
S_{BH} &=& {\rm ln}\, \chi({\cal M}_{Q_0,Q_B})
%\approx 2\pi \sqrt{\ft16\chi(M)(k-\ft12 C_{AB} u^A u^B)}
%\nn\\&=&
\approx 2\pi \sqrt{(D_{ABC} p^A p^B p^C+\ft16\,c_{2A} p^A )(Q_0+\ft{1}{12} D^{AB} Q_A Q_B)}
\label{entropy}
\eeqa
with $\chi({\cal M}_{Q_0,Q_A})$ the moduli space of $Q_0$ instantons
with first Chern class $Q_A$ for an ${\cal N}=4$ SYM living on
the four-cycle $M$ with conjugacy class
$[M]=p^A\alpha_A$ inside the CY.
Formula (\ref{entropy}) perfectly agrees
with the micro/macroscopic M5-brane/supergravity results found in \cite{msw}.

\vskip 1cm
\section*{Acknowledgements}
We thank   M.L. Frau, A. Lerda, M. Mari\~no and U.Bruzzo  for
several useful discussions. Moreover we thank L. Griguolo, D.
Seminara and A. Tanzini for sharing their results with us. R.P.
would like to thank I.N.F.N. for supporting a visit to the
University of Rome II, "Tor Vergata" and the Volkswagen Foundation
of Germany.

\noindent
This work is partially supported by the European Community's Human Potential
Programme under contracts MRTN-CT-2004-512194,
by the INTAS contract 03-51-6346, by the NATO contract NATO-PST-CLG.978785,
and by the Italian MIUR under contract PRIN-2005024045.

\appendix

\section{Factorization algorithm}
\label{fuji}
\setcounter{equation}{0}

%*************************************************************************
%%%%%%%%%%%%%%%%%%%%%%%%%%%% Figura
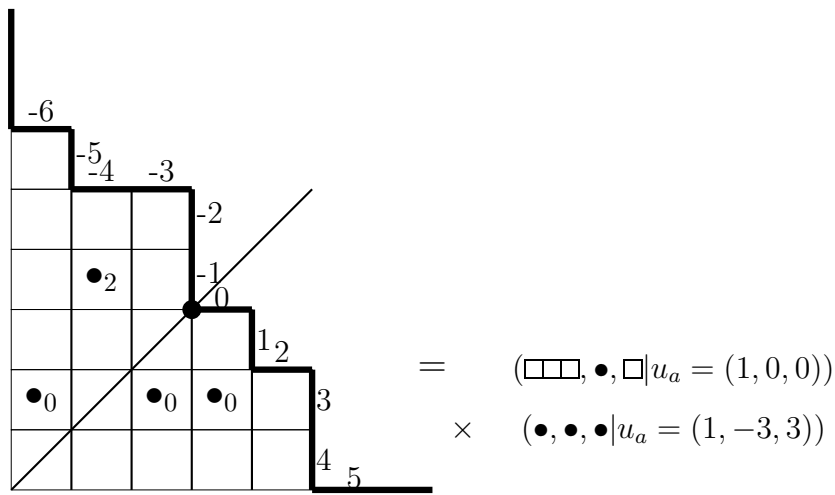
\begin{figure}
\begin{center}
\psset{unit=.8cm}
\begin{pspicture}(0,0)(25,10)
\pspolygon[linewidth=.1pt](1,1)(6,1)(6,2)(1,2)\pspolygon[linewidth=.1pt](1,3)(6,3)(6,2)(1,2)
\pspolygon[linewidth=.1pt](1,3)(5,3)(5,4)(1,4)
\pspolygon[linewidth=.1pt](1,5)(4,5)(4,4)(1,4)\pspolygon[linewidth=.1pt](1,5)(4,5)(4,6)(1,6)
\pspolygon[linewidth=.1pt](1,7)(2,7)(2,6)(1,6)
\psline(2,1)(2,7)\psline(3,1)(3,6)\psline(4,4)(4,1)\psline(5,1)(5,3)
\psline[linewidth=2.4pt](1,7)(1,9)\psline[linewidth=2.4pt](6,1)(8,1)
\psline[linewidth=2.4pt](1,7)(2,7)\psline[linewidth=2.4pt](2,6)(2,7)
\psline[linewidth=2.4pt](2,6)(4,6)\psline[linewidth=2.4pt](4,4)(4,6)\psline[linewidth=2.4pt](4,4)(5,4)
\psline[linewidth=2.4pt](5,3)(5,4)\psline[linewidth=2.4pt](5,3)(6,3)\psline[linewidth=2.4pt](6,1)(6,3)
\rput(8,3){\large =}
\rput(12,3){$\left(\tinyyoung{~~~},\bullet,\tinyyoung{~}\big|u_a=(1,0,0)\right)$}
\rput(8.5,2){$\times$}\rput(12,2){$\left(\bullet,\bullet,\bullet\big|u_a=(1,-3,3)\right)$}
\rput(4.5,4.2){0}\rput(5.2,3.5){1}\rput(5.5,3.3){2}\rput(6.2,2.5){3}\rput(6.2,1.5){4}\rput(6.7,1.2){5}
\rput(4.3,4.6){-1}\rput(4.3,5.6){-2}\rput(2.5,6.3){-4}\rput(3.5,6.3){-3}\rput(2.3,6.6){-5}\rput(1.5,7.3){-6}
\rput(2.5,4.5){\large $\bullet_2$}\rput(1.5,2.5){\large $\bullet_0$}\rput(3.5,2.5){\large $\bullet_0$}
\rput(4.5,2.5){\large $\bullet_0$}
\psline(1,1)(6,6)\psdots[dotscale=2](4,4)
\end{pspicture}
\vskip.5cm
\caption{Factorization algorithm. The diagrams on the right hand side $(Y^*_a,u_a)$
represents the regular (upper) and fractional part (bottom) of tableau $Y$ in the left
hand side. The embedding of the tableaux $Y_a^*$ (in the r.h.s. of the figure)
is given by the boxes marked by $\bullet_a$ in $Y$.}\label{factorfig}
\end{center}
\end{figure}
%%%%%%%%%%%%%%%%%%%%%%%%%%
 The regular and fractional part of a diagram $Y$ can be extracted following
a combinatoric algorithm developed in \cite{Fujii:2005dk}. First,
one associates to the Young tableau $Y$ a sequence $ m_Y(n)$,
$n\in \Z$ made of $0$'s and $1$'s. The sequence is constructed as
follows: to each horizontal (vertical) segment in the Young
tableau profile, we assign a 1 (0). The term $n=0$ in the sequence
is given by the segment to the right of the middle point given by
the intersection between the profile and the main diagonal, see
Fig.\ref{factorfig} where the middle point is marked by a bullet.
$n$ grows from left to right along the Young tableau profile from
$-\infty$ to $\infty$. Notice that a sequence built in this way
satisfy $m_Y(-\infty)=0$ and $m_Y(\infty)=1$. Conversely a
sequence satisfying these boundary conditions specifies uniquely a
Young tableau. For example the Young tableau of
Fig.\ref{factorfig}, representing the partition
$\nu^Y=(5,5,4,3,3,1)$, leads to the sequence $m_Y(n)$ in the first
line in Table \ref{maya}. Next we split the sequence  $m_Y(n)$
into $p$ subsequences $m_{Y_a^*}$ according to the $mod~ p$ parity
of $n$, i.e. \beq m_{Y_a^*} (n)=m_Y( n\, p+a) \eeq The resulting
subsequences are displayed in Table \ref{maya}. Each subsequence
$m_{Y_a^*}$ describes a new Young tableau profile that we denote
by $Y^*_a$, see Table \ref{maya}. In addition the first Chern
classes, $u_a$, can then be read from \beq
u_a=\nu_{a+r-1}-\nu_{a+r}+\delta_{a,r} \label{unu} \eeq with \beq
\nu_a\equiv \#\, \{ m_{Y_a^*}(n)=0\big|n\geq 0\}-\#\, \{
m_{Y_a^*}(n)=1\big|n<0 \} \label{unu2} \eeq We remark that from
this definition $\sum_a \nu_a=0$ since it gives the difference
between the number of horizontal segments to the left and the
number of vertical segments on the right of the middle point.
Therefore giving the $(p-1)$ independent components $u_{a}$'s or
the $\nu_{a}$'s is equivalent.
%%%%%%%%%%%%%%%%%%%%%%%%%%%%%%%%%%%%%%%%%%%%%%%%%%%%%%%%%%%%%%%%%%%%%%%%%%%%%%%%%%%%**

It is interesting to remark that the subdiagrams $Y_a^*$ can be
thought of as embedded inside $Y$. To see this first let us note
that to each invariant box contributing non trivially to the
character (\ref{charinv}) it is associated a hook starting and end
whose on segments labelled by $n_{up}$ and $n_{right}$ satisfying
$n_{up}=n_{right}=a(mod~ p)$ for some $a=0,...p-1$.
%respectively among the edges constituting the profile of $Y$ with
%$n_{up}-n_{right}=0(mod~ p)$.
Let us indicate each invariant box by a bullet with an index $a$
%($a=0,\cdots ,p-1$) if $n_{up}=n_{right}=a(mod~ p)$
(e.g. for the bullet with index $2$ in Fig. \ref{factorfig} we
have $n_{up}=-4=2(mod~3)$, $n_{right}=-1=2(mod~3)$). The bullets
of index $a$ precisely indicate the embedding of $Y_a^*$ inside
$Y$ (see Fig. \ref{factorfig} ). In general the images of $Y_a^*$
in $Y$ overlap, but in the special case when $u_a$ are large
enough
%\nu_0\ll \cdots \ll \nu_{p-1}$
the diagrams $Y_a^*$ become well separated. It is not hard to see
that in such cases the characters (\ref{charinv}) and
(\ref{chinAp}) coincide. Though generally speaking the characters
(\ref{charinv}) and (\ref{chinAp}) are different they give rise to
the same results when $\epsilon_1=-\epsilon_2$ as well as to the
same Poincare' polynomials
%as well as coinciding results for all
%other quantities upon specifying $\epsilon_1=-\epsilon_2$.

For the tableau of Fig.\ref{factorfig} the results of (\ref{unu})
and (\ref{unu2}) are displayed in the last two columns of Table
\ref{maya}.
 Conversely given the set $\{Y_a^*,u_a\}$ one can reconstruct the sequence $m_Y(n)$ and therefore
the tableau $Y$, i.e. the algorithm gives a one-to-one correspondence between a Young tableau $Y$
and the p sets $(Y^*_a,u_a)$ with $\sum_a u_a=1$.
\begin{table}[h]
 \begin{tabular}{|c|ccccccccccccccccc|c|c|c|}
   \hline
   n & -$\infty$     & . & -7 & -6 & -5 & -4 & -3 & -2  & -1 & 0 & 1 & 2& 3 & 4 &  5 &   . & $\infty$ & $Y^*_a$         & $\nu_a$ & $u_a$ \\
   \hline
      $m_Y$ & 0      & 0 & 0  & 1 & 0   & 1  & 1  & 0   & 0  & 1 & 0 & 1& 0 & 0 &  1 &   1 &   1   &       x            &   x     &   x \\
$m_{Y^*_0}$ & 0      &   &    & 1 &     &    & 1  &     &    & 1 &   &  & 0 &   &    &   1 &       &  $\tinyyoung{~~~}$  &   -1    &   1 \\
$m_{Y^*_1}$ & 0      & 0 &    &   & 0   &    &    & 0   &    &   & 0 &  &   & 0 &    &     &   1   &  $\bullet $  &   2    &   -3 \\
$m_{Y^*_2}$ & 0      &   & 0  &   &     & 1  &    &     & 0  &   &   & 1&   &   &  1 &     &       &  $\tinyyoung{~}$  &   -1    &   3 \\
\hline
 \end{tabular}
\caption{\footnotesize The sequences
$m_Y$ or $m_{Y^*_a}$ specify uniquely the Young tableaux $Y$ in fig. \ref{factorfig} and its regular part $
(Y^*_a,u_a=(1,0,0))$. $\nu_a$ or
equivalently $u_a$ describe the fractional part carrying the non-trivial first Chern class . }
 \label{maya}
\end{table}

The regular and fractional part of a diagram can be extracted from $(Y^*_a,u_a)$ by the inverse
algorithm via the identification
\beq
Y_{\rm reg}\leftrightarrow (Y^*_a,u_a=\delta_{a,0}) \quad\quad
Y_{\rm frac} \leftrightarrow (Y^*_a=\bullet,u_a)
\eeq
i.e. from  $(Y^*_a,u_a=\delta_{a,0})$ one reconstructs the subsequences $m_{Y^*_{a,{\rm reg}}}(n)$
and from them the profile given by $m_{Y_{\rm reg}}(n)$ and the tableau $Y_{\rm reg}$.
More precisely, since $u_a=\delta_{a,0}$ implies $\nu_a=0$,  $m_{Y^*_{a,{\rm reg}}}(n)$ are
found by translating the subsequences $m_{Y^*_a}(n)$ in
Table \ref{maya} in such a way that the number of $0$'s for $n\geq 0$ is the same as that of $1$'s
for $n< 0$.

To extract $Y_{\rm frac}$ it is easier.
  $Y_{\rm frac}$ can be found by removing from $Y$ all possible hooks of length a
multiple of $p$.
Indeed according to (\ref{ut}) $u_a$ is invariant under the operation of removing a hook of length
$l p$ or $k_a\to k_a-l$. Therefore the fractional tableau obtained with this operation carries the same
$u_a$ of $Y$.
For the diagram in fig \ref{factorfig} one finds
\beq
\tinyyoung{~,~~~,~\bullet~,~~~~,\bullet~\bullet\bullet~,~~~~~}=\tinyyoung{~,~~~,~~~~~}
\times \tinyyoung{~,~\bullet~~,\bullet\bullet~~\bullet~~}
\eeq
 We have indicated by a bullet the invariant boxes contributing to the character. $Y_{\rm frac}$ is
 the diagram without bullets and is found by removing the length $9$ and length $3$
 hooks containing the bullets in $Y$ (the diagram in the left hand side).
 Notice that both $Y$ and $Y_{\rm reg}$ has as many invariant boxes as boxes in $Y^*_a$.

\section{$\C^2/\Gamma_{p,q}$ toric geometry}
\setcounter{equation}{0} \label{storico}

In this appendix we review the geometry of toric singularities $\C^2/\Gamma_{p,q}$.
We refer the reader to \cite{fulton} for a nice background material.
 A toric variety of complex dimension $d$ is specified by a cone
in $\R^d$ generated by a set of vectors $\vec{v}_a$ (with integer coefficients)
 \beq
 \sigma\equiv \{\quad  \sum_a r_a\, \vec{v}_a \quad \big| \quad  r_a\in \R_+^d \quad \}
 \eeq
  Given $\sigma$, one introduces the dual cone $\sigma^*$, and the
 lattice reductions $\sigma_N,\sigma^*_N$
\beqa
 \sigma^*  & \equiv &  \{\quad \vec{u}\in \R^d\quad \big|
 ~~\vec{u}\cdot \vec{v}\geq 0 \quad \forall\, \vec{v}\in \sigma \quad \}\nn\\
 \sigma_N  & \equiv & \sigma  \cap \Z^d\quad\quad
 \sigma_N^*   \equiv  \sigma^*  \cap \Z^d
 \eeqa
  The lattices $\sigma_N,\sigma^*_N$ encode the basic geometrical data
  of the toric variety. In particular,
 points on $\sigma_N^* $ are in one-to-one correspondence with the set of holomorphic
 functions on the toric variety.  More precisely, let $\{ (a_{1m},\ldots,a_{dm}) \}$ a basis
for $\sigma^*_N$ i.e. the minimal set of vectors in $\sigma^*_N$ such
that any point in $\sigma^*_N$ can be written as
$\sum_{m=1}^D c_m (a_{1m},\ldots, a_{dm})$ with $c_m\in \Z_+$ .
Then the ring of holomorphic
functions on the toric variety can be written as
\beq
\C [\{ x_m \}]/G=
\C [\{ w_1^{a_{1m}}\, w_2^{a_{2m}}\ldots w_d^{a_{dm}} \}]\quad\quad m=1,..D
\eeq
 The variables $x_m$ are clearly not independent, they satisfy $D-d$ relations $G$ that
 can be used to define the variety as an hypersurface on $\C^D$.
 Some basic properties are evident in $\sigma_N$.
 A variety is non-singular if and only if any point inside $\sigma_N$
 can be written as an integer-valued linear combination of $\{ v_a \}$.
Clearly the toric variety can be made regular by adding enough $\vec{v}_a$'s
to the cone $\sigma$. A variety is compact if $\sigma_N$ is isomorphic to $ \Z^d$.

%%%%%%%%%%%%%%%%%%%%%%%%%%%% Figura
\begin{figure}
\begin{center}
\psset{unit=.5cm}
\begin{pspicture}(0,0)(20,10)
\psline[linestyle=dashed,arrows=->](2,0)(2,5)\psline[linestyle=dashed,arrows=->](0,3)(7,3)
\psline[linewidth=1pt]{->}(2,3)(2,4)
\psline[linewidth=1pt]{->}(2,3)(4,0)
\psdots[dotsize=3pt](3,3)(4,3)(5,3)(6,3)(7,3)(2,4)(3,4)(4,4)(5,4)(6,4)(7,4)(2,5)(3,5)(4,5)(5,5)(6,5)(7,5)
\psdots[dotsize=3pt](4,1)(5,1)(6,1)(7,1)(3,2)(4,2)(5,2)(6,2)(7,2)(4,0)(5,0)(6,0)(7,0)
%%%%%%%%%%%%%%%%%%%%%%%%%%%%%%%%%%%%%%%%%%%%%%%%%%%%%%%%%%%%%%%%%%%%%%%%%%%
\psline[linestyle=dashed,arrows=->](12,0)(12,5)\psline[linestyle=dashed,arrows=->](10,3)(17,3)
\psline[linewidth=1pt]{->}(12,3)(13,3)
\psline[linewidth=1pt]{->}(12,3)(14,6)
\psdots[dotsize=3pt](13,3)(13,4)(14,3)(15,3)(16,3)(17,3)(14,4)(14,5)(15,4)(16,4)(17,4)(15,5)(16,5)(17,5)
\rput(4,7){\large $\sigma$}\rput(14,7){\large $\sigma^*$}
\rput(.8,4){(0,1)}\rput(4,-1){(p,-q)}\rput(15,6){(q,p)}\rput(13,2){(1,0)}
\end{pspicture}
\vskip.5cm
\caption{Toric diagram for the $\C^2/\Gamma_{p,q}$ singularity.}\label{torico}
\end{center}
\end{figure}
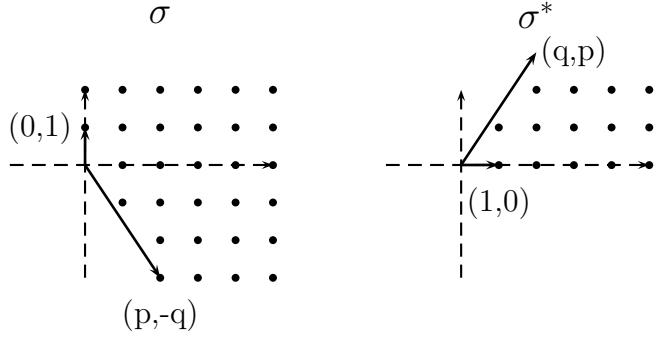
%%%%%%%%%%%%%%%%%%%%%%%%%%
Let us illustrate these abstract notions in the case
of $\C^2/\Gamma_{p,q}$-singularities, $(p,q)$ being coprime numbers with $q<p$.
More precisely $\Gamma_{p,q}$ is a $\Z_p$ action generated by
\beq
\Gamma_{p,q}:~~~~  \left(%
\begin{array}{c}
  z_1 \\
  z_2 \\
\end{array}%
\right) \rightarrow  \left(%
\begin{array}{cc}
  e^{2\pi i q/p} & 0 \\
  0 &   e^{2\pi i/p}\\
\end{array}%
\right)\left(%
\begin{array}{c}
  z_1 \\
  z_2 \\
\end{array}%
\right)
\eeq
 with $z_{1,2}$ the complex coordinates on $\C^2$

The cone $\sigma$ and its dual $\sigma^*$ in this case are generated by the
vectors $\vec{v}_a,\vec{v}_a^*$ given by (see fig \ref{torico})
\beqa
\vec{v}_0&=& (0,1) \quad\quad \vec{v}_{1}=(p,-q) \nn\\
\vec{v}^*_0&=& (1,0) \quad\quad \vec{v}^*_{1}=(q,p)\label{vpq}
\eeqa
 The variables $w_1,w_2$ are built out of invariant combinations of $z_1,z_2$
 \beq
 w_1=z_1^p \quad\quad w_2={z_2\over z_1^q}
 \eeq
%%%%%%%%%%%%%%%%%%%%%%%%%%%% Figura
\begin{figure}
\begin{center}
\psset{unit=.5cm}
\begin{pspicture}(0,0)(20,10)
\psline[linewidth=.2pt,linestyle=dashed,arrows=->](2,3)(2,7)
\psline[linewidth=.2pt,linestyle=dashed,arrows=->](4,3)(7,3)
\psline[linewidth=.2pt,linestyle=dashed](6,1)(8,0)
%\psline[linewidth=.2pt,linestyle=dashed](8,-1)(9.5,-2)
%%%%%   vettori
\psline[linewidth=1.5pt]{->}(2,3)(2,5)
%\rput(1,5){(0,1)}
\psline[linewidth=1.5pt]{->}(2,3)(4,3)
%\rput(4,3.5){(1,0)}
\psline[linewidth=1.5pt]{->}(2,3)(6,1)
%\rput(7.3,2){(2,-1)}
\psline[linewidth=1.5pt]{->}(2,3)(8,-1)
%\rput(8,-.8){(3,-2)}
\psdots[dotsize=2pt](2,7)(4,7)(6,7)(8,7)(2,5)(4,5)(6,5)(8,5)(4,3)(6,3)(8,3)(6,1)(8,1)(8,-1)
%\psdots[dotsize=2pt]
\rput(3.5,4.6){$\sigma_1$}\rput(5.7,2.3){$\sigma_2$}\rput(7,.2){$\sigma_3$}
%%%%%%%%%%%%%%%%%%%%%%%%%%%%%%%%%%%%%%%%%%%%%%%%%%%%%%%%%%%%%%%%%%%%%%%%%%%
\psline[linewidth=1.5pt]{->}(11,3)(12,3)\psline[linewidth=1.5pt]{->}(11,3)(11,4)
%\rput(13,3){(1,0)}\rput(11,5){(0,1)}
\psdots[dotsize=2pt](12,3)(11,4)(13,3)(12,4)(11,5)
\psline[linewidth=1.5pt]{->}(16,3)(16,2)\psline[linewidth=1.5pt]{->}(16,3)(17,6)
\psdots[dotsize=2pt](16,3)(16,2)(17,2)(17,3)(17,4)(17,5)(17,6)(18,2)(18,3)(18,4)(18,5)
\psline[linewidth=1.5pt]{->}(20,3)(22,8)\psline[linewidth=1.5pt]{->}(20,3)(19,0)
\psdots[dotsize=2pt](19,0)(20,2)(20,1)(20,0)(21,5)(21,4)(21,3)(21,2)(21,1)(21,0)(22,7)(22,6)(22,5)(22,4)(22,3)(22,2)(22,1)(22,0)
\psdots[dotsize=3pt](20,3)
\rput(11,8){\large $\sigma^*_1$}\rput(16,8){\large $\sigma^*_2$}\rput(20,8){\large $\sigma^*_3$}
\end{pspicture}
%\vskip.5cm
\caption{Resolution of $\C^2/\Gamma_{3,2}$.}\label{torico1}
\end{center}
\end{figure}
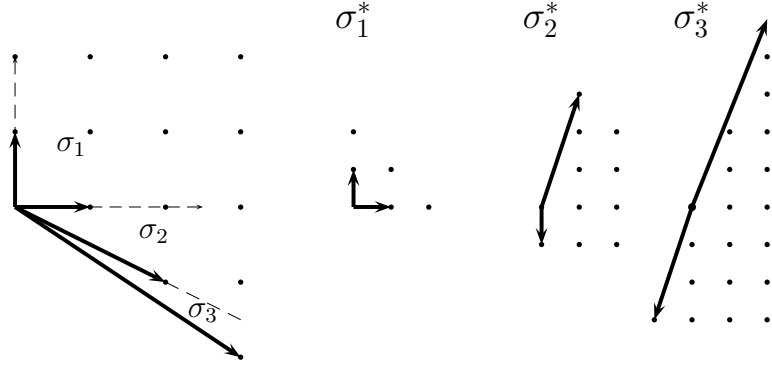
%%%%%%%%%%%%%%%%%%%%%%%%%%
{\bf $A_{p-1}$-singularity}

 Take $q=p-1$. Collecting the basic monomials in $\sigma^*_N$ one finds
\beq
\C [ x_1, x_2, x_3 ]/G=\C [ w_1, w_1^{p-1}\, w_2^p,\,w_1 \, w_2 ]
\eeq
  with
\beq
G: \quad\quad x_1 \, x_2=x_3^p \label{hypzk}
\eeq
 This equation realizes the orbifold as a hypersurface on $\C^3$.

 The variety can be made regular adding
 vectors $v_a$\footnote{We relabel $v_1=(p,-q)$ in (\ref{vpq}) as $v_{p-1}$.}
 \beq
\sigma: \quad\quad  \{ v_a=(a,1-a)\, ,\quad a=0,..p-1 \}
 \eeq
The new cone become the union of $p$-cones $\sigma_{a}$ defined by (see Fig.\ref{torico1})
 \beqa
 \sigma_{a} &=& \{ (a,1-a), (a+1,-a) \}     \quad\quad  a=0,\ldots,p \nn\\
\sigma^*_{a} &=& \{ (1-a,-a), (a,a+1) \}
 \eeqa
  This corresponds to blow up $(p-1)$ $\PP_1$'s one for each extra $v_a$.
The cone $\sigma^*_N$ is made out of $p$ cones $\sigma^*_{N,a}$
with polynomial ring \beq \sigma^*_{N}:\quad\quad  \oplus_{a}\,
\C[w_1^{1-a}\, w_2^{-a},w_1^{a}\, w_2^{a+1} ]= \oplus_{a}\,
\C[z_1^{p-a}\, z_2^{-a},z_1^{a+1-p}\, z_2^{a+1} ] \eeq with \beq
 w_1=z_1^p \quad\quad w_2={z_2\over z_1^{p-1}}
 \eeq
  This is precisely the result we found for the $A_{p-1}$-polynomial ring in (\ref{chiralap}).

{\bf $O_{\PP_1}(-p)$-singularity}

 Take $q=1$. %Collecting the basic monomials in $\sigma^*_N$ one finds
%\beq
%\C [ x_0,x_1,\ldots,x_p ]/G=\C [ w_1, w_1 w_2,\ldots,w_1\,w_2^{p} ]
%\eeq
% with $x_a$ satisfy the relations
% \beq
% G\quad\quad x_a\, x_{n-a}=x_b \, x_{n-b}   \quad\quad \forall \, n,a,b
% \eeq
%  The variety can be made regular adding a single vector $v_1=(1,0)$
% to the cone $\sigma$.
 The resolved variety  $O_{\PP_1}(-p)$ is described by the union of $2$-cones
 $\sigma_{1}$ and $\sigma_{2}$ defined by
 \beqa
 \sigma_{1} &=& \{ (0,1), (1,0) \}     \quad\quad
  \sigma_{2} = \{ (1,0), (p,-1) \}    \nn\\
\sigma^*_{1} &=& \{ (1,0), (0,1) \} \quad\quad
  \sigma^*_{2} = \{ (0,-1), (1,p) \}
 \eeqa
  This corresponds to blow up of a $\PP_1$'s corresponding to the
 extra $v_2=(0,1)$.
The cone $\sigma^*_N$ is made out of $2$ cones with polynomial
ring \beqa \C[w_1, w_2 ]\oplus \C[w_2, w_1\,w_2^p ]=\C[z_1^{p} ,
{z_2\over z_1} ]\oplus \C[{z_1\over z_2} , z_2^p ] \eeqa and \beq
 w_1=z_1^p \quad\quad w_2={z_2\over z_1}
 \eeq
  This is precisely the result we found in (\ref{chiralop}) for the $O_{\PP_1}(-p)$ polynomial ring.

{\bf $\C^2/\Gamma_{5,2}$-singularity}

 The cone associated to the resolved variety $\widehat{\C^2/\Gamma_{5,2}}$ is made out of three cones
 \beqa
 \sigma_{1} &=& \{ (0,1), (1,0) \}     \quad\quad
  \sigma_{2} = \{ (1,0), (3,-1) \}  \quad\quad  \sigma_{3} = \{ (3,-1), (5,-2) \}  \nn\\
  \sigma^*_{1} &=& \{ (0,1), (1,0) \}     \quad\quad
  \sigma^*_{2} = \{ (0,-1), (1,3) \}  \quad\quad  \sigma^*_{3} = \{ (-1,-3), (2,5) \}  \nn\\
 \eeqa
  This corresponds to blow up of two $\PP_1$'s corresponding to the
 extra $v_2=(0,1)$ and $v_3=(3,-1)$.
The cone $\sigma^*_N$ is made out of $3$ cones with polynomial
ring \beqa \C[w_1, w_2 ]\oplus \C[w_2^{-1}, w_1\,w_2^3
]+\C[w_1^{-1}\,w_2^{-3}, w_1^2\,w_2^5 ] =\C[z_1^{5} , {z_2\over
z_1^2} ]\oplus \C[{z_1^2\over z_2} , {z_2^3 \over z_1} ]\oplus
\C[{z_1\over z_2^3} , z_2^5 ]\nn \eeqa and \beq
 w_1=z_1^5 \quad\quad w_2={z_2\over z_1^2}
 \eeq
  This is precisely the result we found in (\ref{chiral52}) for the polynomial ring.

%\bibliographystyle{JHEP-2}
%\bibliographystyle{amsplain}
%\bibliography{ref}

\providecommand{\href}[2]{#2}\begingroup\raggedright\endgroup

\end{document}